\documentclass[%
reprint,
superscriptaddress,
 amsmath,amssymb,
 aps,
prb,
]{revtex4-2}

\usepackage{graphicx}
\usepackage{dcolumn}
\usepackage{bm}
\usepackage{hyperref}
\usepackage{epsfig}
\usepackage{float}

\usepackage{color}
\usepackage{xcolor}

\renewcommand{\H}{\hat{\mathcal{H}}}
\newcommand{\hc}{\text{h.\,c.}}
\newcommand{\Zt}{$\mathbb{Z}_2$ }
\renewcommand{\ij}{\langle i, j \rangle}
\newcommand{\ad}{\hat{a}^{\dagger}}
\renewcommand{\a}{\hat{a}}
\newcommand{\tauZ}{\hat{\tau}^{z}}
\newcommand{\tauX}{\hat{\tau}^{x}}

\begin{document}

\title{Confinement Induced Frustration in a One-Dimensional $\mathbf{\mathbb{Z}_2}$ Lattice Gauge Theory}

\author{Matja\v{z} Kebri\v{c}}
\affiliation{Department of Physics and Arnold Sommerfeld Center for Theoretical Physics (ASC), Ludwig-Maximilians-Universit\"at M\"unchen, Theresienstr. 37, M\"unchen D-80333, Germany}
\affiliation{Munich Center for Quantum Science and Technology (MCQST), Schellingstr. 4, D-80799 M\"unchen, Germany}

\author{Umberto Borla}
\affiliation{Munich Center for Quantum Science and Technology (MCQST), Schellingstr. 4, D-80799 M\"unchen, Germany}
\affiliation{Physik-Department, Technische Universit\"at M\"unchen, 85748 Garching, Germany}

\author{Ulrich Schollw\"ock}
\affiliation{Department of Physics and Arnold Sommerfeld Center for Theoretical Physics (ASC), Ludwig-Maximilians-Universit\"at M\"unchen, Theresienstr. 37, M\"unchen D-80333, Germany}
\affiliation{Munich Center for Quantum Science and Technology (MCQST), Schellingstr. 4, D-80799 M\"unchen, Germany}

\author{Sergej Moroz}
\affiliation{Munich Center for Quantum Science and Technology (MCQST), Schellingstr. 4, D-80799 M\"unchen, Germany}
\affiliation{Physik-Department, Technische Universit\"at M\"unchen, 85748 Garching, Germany}
\affiliation{Department of Engineering and Physics, Karlstad University, Karlstad, Sweden}

\author{Luca Barbiero}
\affiliation{Institute for Condensed Matter Physics and Complex Systems,
DISAT, Politecnico di Torino, I-10129 Torino, Italy}

\author{Fabian Grusdt}
\email{Fabian.Grusdt@physik.uni-muenchen.de}
\affiliation{Department of Physics and Arnold Sommerfeld Center for Theoretical Physics (ASC), Ludwig-Maximilians-Universit\"at M\"unchen, Theresienstr. 37, M\"unchen D-80333, Germany}
\affiliation{Munich Center for Quantum Science and Technology (MCQST), Schellingstr. 4, D-80799 M\"unchen, Germany}

\date{\today}

\begin{abstract}
Coupling dynamical charges to gauge fields can result in highly non-local interactions with a linear confining potential. As a consequence, individual particles bind into mesons which, in one dimension, become the new constituents of emergent Luttinger liquids. Furthermore, at commensurate fillings, different Mott-insulating states can be stabilized by including nearest-neighbour (NN) interactions among charges. However, rich phase diagrams expected in such models have not been fully explored and still lack comprehensive theoretical explanation. Here, by combining numerical and analytical tools, we study a simple one-dimensional $\mathbb{Z}_2$ lattice gauge theory at half-filling, where U$(1)$ matter is coupled to gauge fields and interacts through NN repulsion. We uncover a rich phase diagram where the local NN interaction stabilizes a Mott state of individual charges (or partons) on the one hand, and a Luttinger liquid of confined mesons on the other. Furthermore, at the interface between these two phases, we uncover a highly frustrated regime arising due to the competition between the local NN repulsion and the non-local confining interactions, realizing a pre-formed parton plasma. Our work is motivated by the recent progress in ultracold atom experiments, where such simple model could be readily implemented. For this reason we calculate the static structure factor which we propose as a simple probe to explore the phase diagram in an experimental setup.
\end{abstract}

\maketitle


\section{Introduction}

Recently, quantum simulation based on cold atomic gases has allowed to make significant progress in the exploration of many-body quantum physics. This has opened a new venue for the study of lattice gauge theories \cite{Wilson1974, Kogut1979} where numerical calculations are notoriously difficult in many situations, especially when doping is finite.
For instance, we have seen remarkable success in implementing U$(1)$ lattice gauge theory (LGT) models with tunable parameters \cite{Yang2020, Mil2020, Zhou2021, Aidelsburger2021, Martinez2016}.
Significant progress has also been made in simulating \Zt LGT models by using Floquet schemes \cite{Barbiero2019}, with experimentally realized building blocks which have potential to be scaled to bigger system sizes \cite{Goerg2019, Schweizer2019}. Furthermore, it has also been shown that the \Zt LGTs could be implemented using superconducting qubits \cite{Zohar2017, HomeierPRB2021}.
More recently, a Rydberg tweezer array implementation \cite{Homeier2022} which utilizes the so called local-pseudo-generators \cite{Halimeh2021LPG, Halimeh2021DisorderFree} has been proposed.
In addition, digital schemes have been put forward \cite{Irmejs2022, Armon2021, Greenberg2022, Zohar2017}, and during the time of writing this manuscript, a realization of a \Zt LGT Hamiltonian was achieved on a digital quantum computer \cite{Mildenberger2022}.
All these accomplishments have paved the way towards successful experimental simulation of \Zt LGTs and show that tremendous advancement has been made in this field, which has long been limited to mostly theoretical considerations \cite{Schweizer2019}.

Here we explore a phase diagram of a realistic \Zt LGT model where U$(1)$ matter is coupled to a \Zt gauge field, by using analytical arguments and large-scale DMRG simulations \cite{Schollwoeck2011, White1992}. We use the \textsc{SyTen} toolkit \cite{hubig:_syten_toolk, hubig17:_symmet_protec_tensor_network} for finite DMRG calculations and TeNPy \cite{tenpy} for infinite DMRG (iDMRG) calculations. The system features an interesting interplay of bare nearest-neighbor (NN) interactions of the conserved and mobile \Zt charges (partons) on one hand, and non-local interactions mediated by the gauge field on the other hand, which induce linear confinfining potential of particle pairs. We focus on the filling of one-half, where we reveal a parton-plasma like phase in a strongly frustrated regime, where the two types of interactions have opposite effects.

Previous theoretical studies of the system we consider focused on liquid phases at generic fillings and without the NN repulsion \cite{Borla2020PRL}, near-full filling \cite{Das2021}, and it has been shown that the NN repulsion leads to mesonic Mott insulating states at two-thirds filling \cite{Kebric2021}. At this commensurate filling, $n = 2/3$, the gauge-mediated confining potential and the NN repulsion act in concert without creating frustration, which results in a phase diagram with three different regimes: deconfined partons in the absence of the confining field, dimers (mesons) consisting of pairs of confined partons \cite{Borla2020PRL}, and a Mott state of crystallized dimers \cite{Kebric2021}. The model is in fact confining for any nonzero value of the \Zt electric field term, as shown in Refs.~\cite{Borla2020PRL,Kebric2021}. 
Interestingly, the deconfined and confined partons both exhibit Luttinger liquid behaviour in the absence of the NN interactions or for generic fillings, with gapless deconfined collective excitations. However, the constituents of such Luttinger liquid drastically change when the partons become confined, which is reflected in a change of Fermi momentum and can be detected by measuring Friedel oscillations at a sharp edge of the system \cite{Borla2020PRL}. 

In this work we study the case of half-filling $n = 1/2$, where the NN repulsion and the confining force have opposite effects, thus giving rise to quantum frustration. In the absence of the gauge-mediated confining potential we obtain deconfined partons, which can crystallize into a Mott state by applying strong enough repulsive NN interaction. 
On the other hand, partons become confined for any non-zero value of the gauge mediated confining potential, thus forming a Luttinger liquid (LL) of confined mesons. Hence the Mott state is destabilized by the increasing value of the confining \Zt electric field term. At the transition from the Mott regime to the LL regime the system becomes highly frustrated and realizes a parton-plasma behaviour on short-to-intermediate length scales. There the parton correlation length exceeds the typical distance between partons, which results in an apparent screening of the confinement potential on short length scales. However the partons still remain confined, which is reflected in the behaviour on longer length scales.
These aforementioned regimes at half-filling are summarized in the phase diagram in Fig.~\ref{fig:DiagramSketch}.

The model we discuss in this work can be simulated experimentally by the setups mentioned in the beginning. A particularly suitable approach would be to use the Rydberg tweezer setup with local-pseudo-generators \cite{Halimeh2021DisorderFree, Halimeh2021LPG}. We propose the static structure factor as a simple probe which could be used in such experiments, since it could be readily obtained from snapshots. Recent developments in advanced state tomography reconstruction schemes \cite{Lange2022} could also be used to gain even better insights into our model within existing experiments.

This article is organized as follows: we start with the introduction of the model.
We continue with analytical considerations of different tractable analytical limits. To do so we use different mappings of the original model to more descriptive effective models.
In the next section we complement the analytical considerations with state of the art DMRG calculations and map out the complete phase diagram at half-filling.
Next, we discuss a peculiar regime of degenerate dimer states which correspond to maximum frustration. To this end we consider the parton Green's function and the pair-pair correlation function.
Finally we draw conclusions and offer an outlook in the last section.

\section{Model}\label{ModelIntro}

\begin{figure}[t]
\centering
\epsfig{file=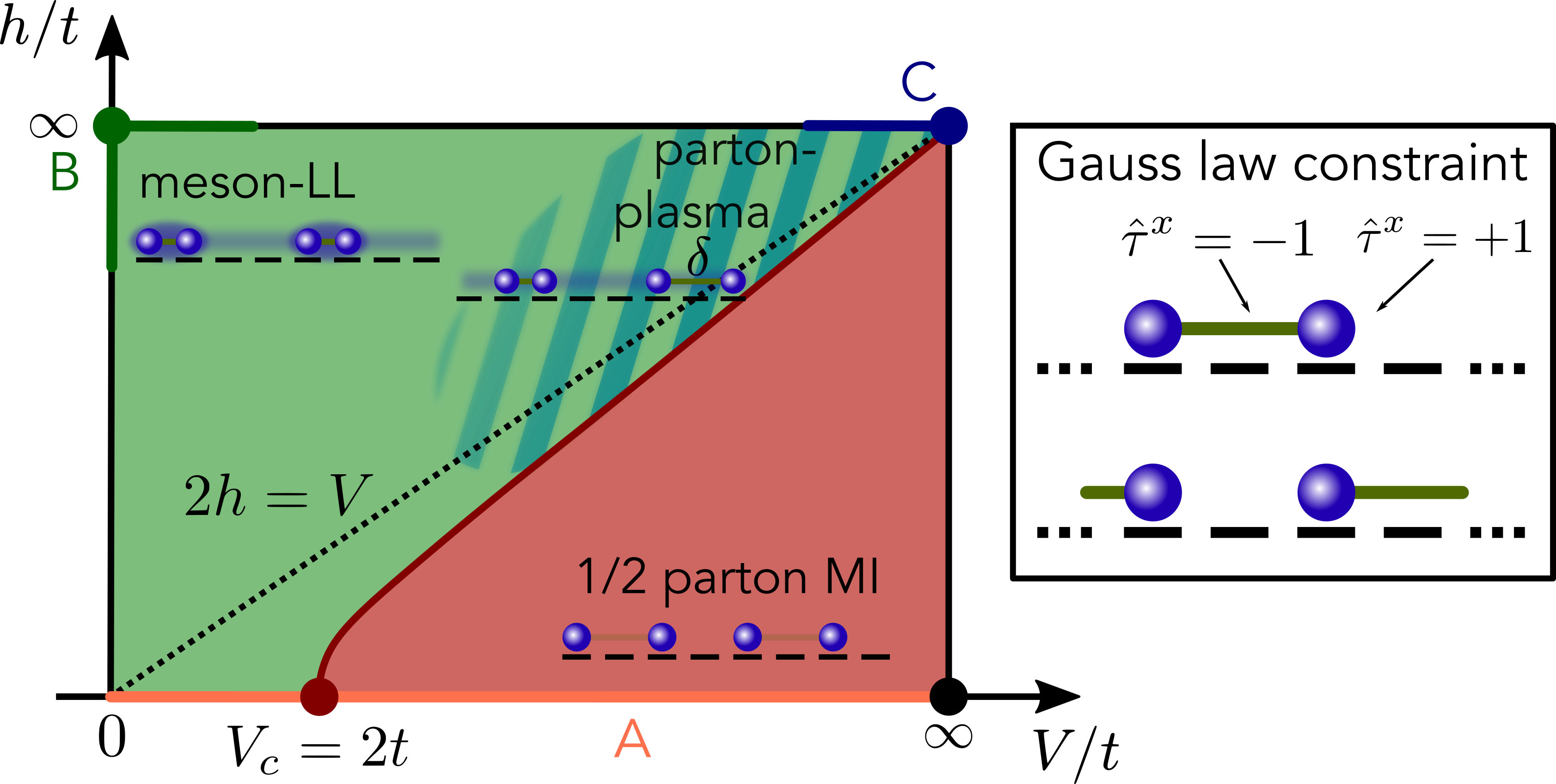, width=0.48\textwidth}
\caption{A sketch of the phase diagram of the \Zt LGT model \eqref{eq:LGT_Model} at half-filling together with the Gauss law constraint. The tractable analytical limits of our model are: (A) hard-core bosons with NN interaction $h = 0, V \geq 0$; (B) tightly confined dimers $h \rightarrow \infty$; and (C) fluctuating dimers $h,V \gg t$. From these limits we can identify the region of confined meson Luttinger liquid (LL) (green), the Mott state consisting of individual partons (red), and the pre-formed parton-plasma region along the $2h = V$ line (blue stripes).}
\label{fig:DiagramSketch}
\end{figure}

We consider a \Zt lattice gauge theory with nearest-neighbor (NN) interactions among $\mathbb{Z}_2-$charged U$(1)$ matter particles \cite{Borla2020PRL, Kebric2021}
\begin{equation}
    \H = -t \sum_{\ij} \left ( \ad_{i} \tauZ_{\ij} \a_{j} + \hc \right ) - h \sum_{\ij} \tauX_{\ij} + V \sum_{\ij} \hat{n}_i \hat{n}_j.
    \label{eq:LGT_Model}
\end{equation}
Here $\ad$ is the hard-core boson creation operator and $\hat{\tau}^{\mu}, \mu \in \{ x, z \}$ are Pauli matrices representing the \Zt gauge ($\mu = z$) and electric ($\mu = x$) fields defined on the links between lattice sites. In addition we consider the Gauss law \cite{Prosko2017}
\begin{equation}
    \hat{G}_j = (-1)^{\hat{n}_j} \tauX_{j-1, j} \tauX_{j, j+1} = \pm 1,
    \label{eq:GaussLaw}
\end{equation}
which divides our Hilbert space into different physical sectors. We chose the sector where $\hat{G}_j = 1, \forall j$. Such construction ensures that the \Zt electric field has to change its sign across an occupied lattice site which gives us a convenient physical interpretation, where we consider the pairs of particles to be connected with \Zt electric strings ($\tauX = -1$) and anti-strings ($\tauX = +1$), see Fig.~\ref{fig:DiagramSketch}. These reflect the orientation of the \Zt electric field. Hence the first term in the Hamiltonian \eqref{eq:LGT_Model} is the usual hopping term with an additional \Zt gauge operator $\tauZ$, which ensures the string attachment to the particle. The second term introduces a linear confining potential of the pairs with string tension $2h$, and the third term is a NN interaction between the hard-core bosons.

In the absence of the electric field term $h$ and NN interaction $V$, the system can be mapped into a free fermion model via an extension of the Jordan-Wigner transformation \cite{Prosko2017}. For nonzero values of the \Zt electric field the model exhibits confinement of the particle pairs \cite{Borla2020PRL} which is a consequence of the translational symmetry breaking in the so called string-length basis \cite{Kebric2021}.

Throughout this paper we will focus on a half-filled chain, $n = 1/2$ unless stated otherwise.

\section{Analytical Limits}\label{AnalyticalLims}

We first discuss the 1D \Zt LGT model \eqref{eq:LGT_Model} in three tractable limits highlighted in Fig.~\ref{fig:DiagramSketch}. This will allow us to identify individual regions in the phase diagram and help us benchmark the numerical results in the next section.

\subsection{Zero electric field term}\label{AnalyticalLims_ZeroField}
The original model \eqref{eq:LGT_Model} reduces to a hard-core bosonic model with NN repulsion in the absence of the \Zt electric field term ${h = 0, V \geq 0}$, see Fig.~\ref{fig:DiagramSketch} (A). Absence of the electric field term means that there are no fluctuations in the \Zt gauge field (no string tension), which can thus be eliminated by attaching the \Zt strings to charge operators \cite{Borla2020PRL} reminiscent of the Jordan-Wigner transformation \cite{Giamarchi2004}. Hence the model can be rewritten as
\begin{equation}
    \H_c = - t \sum_{\langle i, j\rangle}
    \left ( \hat{c}^{\dagger}_i  \hat{c}_j + \hc 
    \right )
    + V \sum_{\langle i, j\rangle} \hat{n}_i \hat{n}_j,
    \label{eq:SLFermionicHamiltonian}
\end{equation}
where $\hat{c}^{\dagger}_i (\hat{c}_j)$ is the dressed hard-core boson creation (annihilation) operator. Note that we obtain a free particle model when also $V = 0$.

The bosonic hard-core model with NN interactions, Eq.~\eqref{eq:SLFermionicHamiltonian}, can be mapped into a $XXZ$ spin-$1/2$ chain \cite{Giamarchi2004}. Such model undergoes a transition to the antiferromagnetic state for half-filling at $V = 2t$, which corresponds to a gapped Mott state of the original model \cite{Giamarchi2004}. Hence our original model \eqref{eq:LGT_Model} undergoes a transition to a Mott state when $h = 0$ and $V > 2t$, (see also Appendix~\ref{SuppXXZmodel} for details).

\subsection{Tightly confined dimers}\label{AnalyticalLims_ConfinedDimers}
Next, we consider the limit where the \Zt electric field term is much greater than the other parameters in our model $h \gg t, V$. In such regime the dimers are tightly confined and can be considered as individual hard-core bosons. Such mapping can be performed e.g., via the so-called squeeze construction, where one defines the new hard-core boson as
$\hat{d}_j^{\dagger} = \a_j^{\dagger} \tauZ_{j, j+1} \a_{j+1}^{\dagger}$ \cite{Borla2020PRL}.
By performing second order perturbation theory, where hopping $t$ is considered to be a small perturbation, the original model \eqref{eq:LGT_Model} can be mapped to hard-core bosons with NN interaction \cite{Borla2020PRL}
\begin{equation}
    \H_d = - \tilde{t} \sum_{\langle i, j\rangle}
    \left ( \hat{d}^{\dagger}_i  \hat{d}_j + \hc 
    \right )
    + \tilde{V} \sum_{\langle i, j\rangle} \hat{n}_i \hat{n}_j,
    \label{eq:SLFermionicHamiltonianDimers}
\end{equation}
where $\hat{n}_{j} = \hat{d}^{\dagger}_{j} \hat{d}_{j}$, $\tilde{t} = \frac{t^2}{2h}$ and ${\tilde{V} = 2\tilde{t} + V}$. Note that we included the NN interactions which can be considered as small perturbation since $V \ll h$. An important detail in such mapping is also the fact, that the effective filling is modified since a pair of particles on two lattice sites is squeezed into one single particle on one lattice site. Hence the filling changes as $n_d = \frac{1}{\frac{2}{n}-1}$, where $n_d$ is the filling in the effective model \eqref{eq:SLFermionicHamiltonianDimers} \cite{Kebric2021}.

This is effectively the same Hamiltonian as in the case where the string tension is zero and we only consider NN interactions in the original model, which we discussed in the previous Section \ref{AnalyticalLims_ZeroField}. However, the effective particles considered here are tightly confined dimers, or mesons, which is in stark contrast to the individual partons considered in model Eq.~\eqref{eq:SLFermionicHamiltonian}.

The effective NN interaction in such model can be tuned by modifying the value of $V$. Although one is limited to the regime where $V \ll h$, the effective NN interaction strength $\tilde{V}$ can nevertheless be tuned to rather strong values in comparison to hopping $\tilde{t}$. In fact, any nonzero value of the NN repulsion $V > 0$ results in $\tilde{V} > 2 \tilde{t}$, while staying in the limit where $t,V \ll h$, since $\tilde{V} = 2 \tilde{t} + V$.

The effective model can also be mapped to the XXZ spin-$1/2$ chain in the same way as the hard-core bosonic model in Sec.~\ref{AnalyticalLims_ZeroField}. As already discussed before, the XXZ spin-$1/2$ chain undergoes a transition to the gapped antiferromagnetic state when $\tilde{V} > 2\tilde{t}$, \emph{and} when the magnetization of the chain is zero, i.e., when the effective filling of the hard-core bosonic model equals to $n_d = 1/2$ \cite{Giamarchi2004}. However, the system remains a Luttinger liquid of confined dimers for half-filling in the original model, since this corresponds to $n_d = 1/3$ in the effective model.
Hence we have a Luttinger liquid of confined dimers for $h \gg V,t$ which is depicted as region (B) in Fig.~\ref{fig:DiagramSketch}.

\begin{figure}[t]
\centering
\epsfig{file=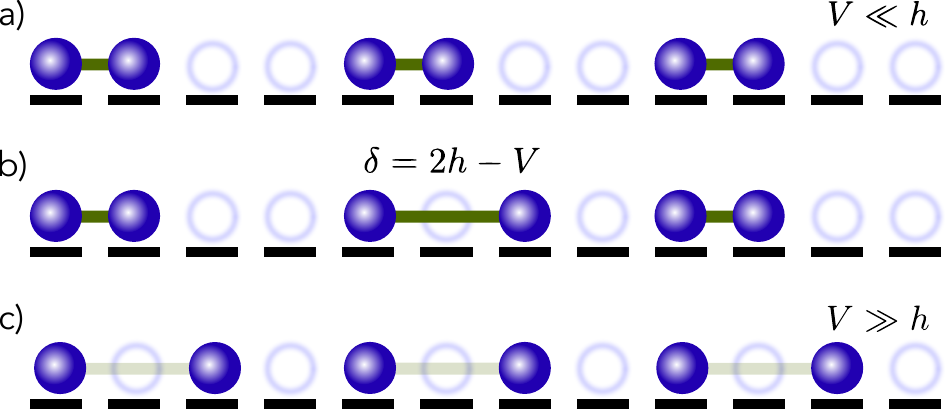, width=0.48\textwidth}
\caption{a) Tightly confined dimers $l = 0$ which are energetically favorable when $2h>V$. b) Energy cost of extending the tightly confined dimer to an extended dimer $l = 1$. c) Extended dimer configuration when $V>2h$.}
\label{fig:DimerConfigurations}
\end{figure}

\begin{figure*}[t!]
\centering
\epsfig{file=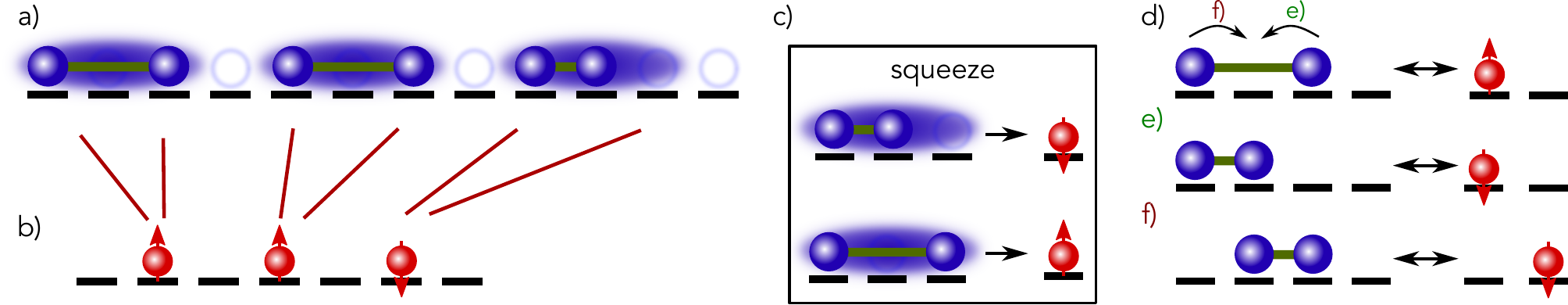, width=0.98\textwidth}
\caption{Mapping of the \Zt LGT model a) to an effective spinful model b). c) The definition of the squeezed space construction for $2h = V$ and $V, h \gg t$.
d) Extended dimer $ l = 1$ where the bent arrows represent hypothetically allowed hoppings of the particles in the dimer.
e) Configuration when the right particle in d) hops to the left. f) Configuration when the left particle in d) hops to the right. Corresponding configurations in the squeezed space are depicted on the right, where the red spheres with arrows represent the pseudospin-up and pseudospin-down respectively.}
\label{fig:DimerSqueeze}
\end{figure*}

\subsection{Fluctuating dimers}\label{AnalyticalFluctuating}
Finally we consider a regime where the \Zt electric field term and the NN repulsion are comparable, but much greater than the hopping parameter $V \approx 2h \gg t$.

\emph{Absence of hopping.--}
We start by first considering a very simple limit where we set $t = 0$. Here the interplay of the confining electric field term and the NN interaction results in two different types of dimers which are energetically favorable for different values of the parameters $h$ and $V$. We first note that when a dimer gets longer by one lattice site the energy cost associated with the electric field is $\Delta E_{+1} = +2h$. Hence we would expect that the dimers tend to be as short as possible, meaning that we have tightly confined pairs where particles are sitting next to each other as depicted in Fig.~\ref{fig:DimerConfigurations} a).
However, since we also have the NN repulsion a tightly confined dimer, where particles are sitting next to each other, will have an extra energy cost of $\Delta E_{NN} = +V$. Hence the energy of an isolated dimer in a regime where $h,V >0$ can be written as
\begin{equation}
    E_{l} = V \delta_{0, l} + 2h(l+1),
    \label{eq:DimerEnergy}
\end{equation}
where $l$ is the number of empty sites between particles in a dimer and $\delta_{0, l}$ is the Kronecker delta function. By considering the energy difference between dimers of length $l=1$ and $l=0$ we define
\begin{equation}
    \delta = E_{1} - E_{0} = 2h - V.
    \label{eq:DimerEnergyDifference}
\end{equation}

We thus obtain two different types of meson configurations: the \emph{tightly confined} dimer $l=0$, will be favorable when $2h>V$ and the \emph{extended} dimer $l=1$, will be favorable when $V > 2h$, see Fig.~\ref{fig:DimerConfigurations} b). Longer dimers ($l>1$) will always be unfavorable since any additional link between confined partons costs additional energy $2h$. We therefore obtain a clear boundary between two different regimes on the line $2h = V$. For $2h>V$ we expect tightly confined dimers, which at half-filling can be placed rather freely on the lattice. This can be done as long as there is at least one free site between adjacent dimers. As a result we get an extensive degeneracy and we expect a Luttinger liquid like behaviour of confined dimers when $t$ is added as perturbation. On the other hand when $V > 2h$ we get a simple Mott state of partons, see Fig.~\ref{fig:DimerConfigurations} c), which only has a $2-$fold overall degeneracy.

\begin{figure*}[t!]
\centering
\epsfig{file=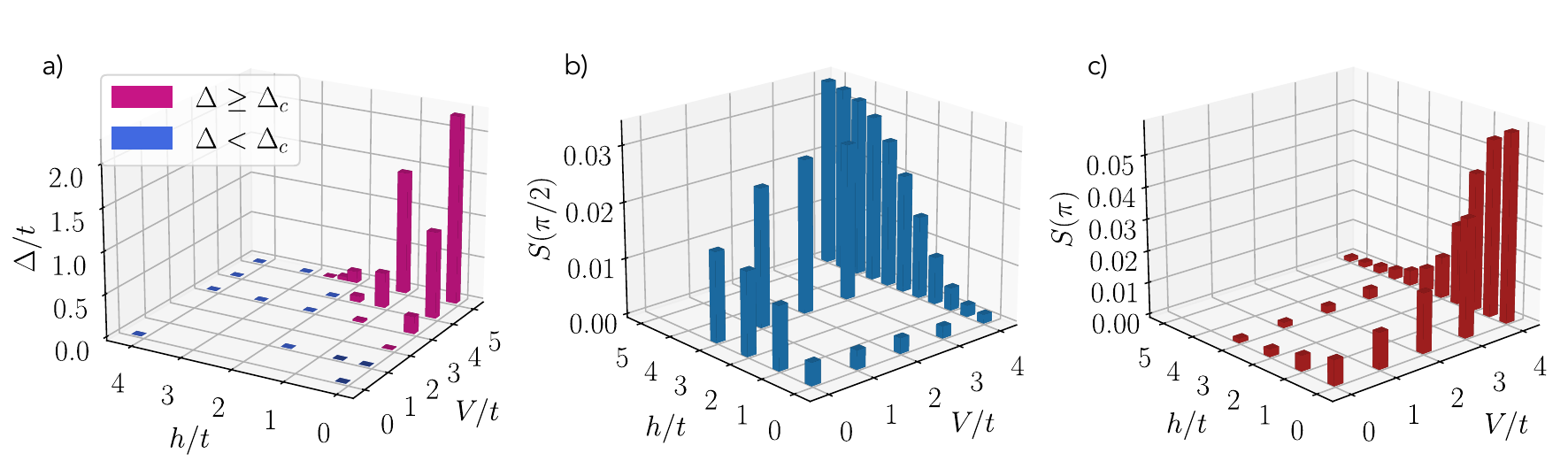, width=0.98\textwidth}
\caption{a) Phase diagram of the charge gap extrapolated in the thermodynamic limit as a function of $h$ and $V$. A substantial charge gap develops for $V > 2t$ which is again destabilised for $2h > V$. The cutoff for the color scale was chosen to be at $\Delta_c = \Delta(h/t = 0, V/t = 2)$. In b) and c) we show the height of the peaks of the structure factor Eq.~\eqref{eq:StaticStructureFactor} for different momenta $k \approx \pi/2$ b) and $k \approx \pi$ c).}
\label{fig:PhaseDiagram}
\end{figure*}

\emph{Hopping as perturbation.--}
Next, we include hopping which we consider to be weak in comparison to the other two parameters $h,V \gg t$. Having strong \Zt electric field terms $h$, means that long dimers are energetically penalized and we restrict our Hilbert space to dimers of length $l<2$. This means that we only consider the \textit{tightly confined} dimers ($l=0$), where the particles are nearest neighbours, and the \textit{extended} dimers ($l=1$), where there is one empty lattice site between two particles which make up the dimer. Such constraint allows us to squeeze the two types of dimers into two different types of mesons which we can consider to be of pseudospin-up (\textit{extended} dimers) or pseudospin-down (\textit{tightly confined} dimers) type, see Fig.~\ref{fig:DimerSqueeze}~a)-c). Note that we always squeeze three lattice sites into one site in the effective model. It does not matter if the dimer is short or extended, see Fig.~\ref{fig:DimerSqueeze} c). Our new spinful particles also have a hard-core property since having two pseudospins on the same lattice site does not represent any physical configuration.

Combining all of these considerations allows us to write a new effective Hamiltonian for the squeezed particles
\begin{multline}
 	\H_s = \hat{P} \left[ - t \sum_{j} \left ( \hat{c}^{\dagger}_{\uparrow, j}  \hat{c}_{\downarrow, j+1} + \hat{c}^{\dagger}_{\uparrow, j}  \hat{c}_{\downarrow, j} + \hc \right ) \right. \\
 	\left. + \frac{t^2}{2(\delta+V)} \sum_{j} \hat{n}_{\downarrow, j} \hat{n}_{\uparrow, j+1} \right. \\
 	\left. + \frac{t^2}{\delta+V} \sum_{j} \hat{n}_{\downarrow, j} \hat{n}_{\downarrow, j+1}  + \delta \sum_{j} n_{\uparrow, j} \right] \hat{P},
	\label{eq:EffectiveSpinfulwithField}
\end{multline}
where $\hat{c}^{\dagger}_{\uparrow}$ is the pseudospin-up creation operator, $\hat{c}^{\dagger}_{\downarrow}$ is the pseudospin-down creation operator, $\hat{n}_{\downarrow, j} = \hat{c}^{\dagger}_{\downarrow, j}
\hat{c}_{\downarrow, j}$, and 
$\hat{n}_{\uparrow, j} = \hat{c}^{\dagger}_{\uparrow, j} \hat{c}_{\uparrow, j}$.
In addition, $\hat{P}$ is a projector to a subspace with no double occupancy, and into a subspace excluding configurations which are highly energetically penalized due to the NN repulsion, due to the fact that ${V \gg t}$. More precisely, the configurations which we project out are those where an up-particle is a left neighbour to any other particle since no NN configurations of particles from different dimers are allowed (see Appendix~\ref{FluctDimers} for details). The first terms in the Hamiltonian come directly from the hopping of the individual particles in the dimer. In the pseudospin language this results in a spin-flip plus hopping to a nearest neighbour term and an on-site spin-flip term, see Fig.~\ref{fig:DimerSqueeze} d)-f). These are all direct consequences of our construction and the restriction of dimers to be shorter than $l<2$ (see Appendix~\ref{FluctDimers}). The NN terms are second order processes and come from restricted virtual hopping of the spins when two particles are sitting next to each other. The last term is the energy contribution $\delta$, which sets in when we are slightly away from the $2h = V$ regime and comes directly from Eq.~\eqref{eq:DimerEnergyDifference}.

From the considerations written above we thus expect strong fluctuations between tightly confined and extended dimers close to $2h = V$. Such fluctuations are interesting in the highly doped regime which is exactly the case at half-filling. Although the particles are overall confined, strong fluctuations in the lengths of such dimers between $l = 0$ and $l = 1$ are energetically allowed. This means that such particles effectively do not feel the linear confinement on short length scales comparable to their average spacing. In other words we expect short-range plasma-like fluctuations of partons in this regime.

On the other hand, the situation becomes clearer when one of the parameters becomes dominant. For example when $2h \gg V$, we regain the Luttinger liquid of tightly confined dimers as discussed in Sec.~\ref{AnalyticalLims_ConfinedDimers} since $\delta > 0$ would penalize the extended dimers. Conversely, when $2h \ll V$ we get a Mott insulating state of individual partons $\hat{a}$, as discussed in Sec.~\ref{AnalyticalLims_ZeroField}.

\subsection{Summary of the analytical results}
We can now distinguish different regions in the phase diagram at half-filling as a function of $h$ and $V$ which we sketch in Fig.~\ref{fig:DiagramSketch}. In the absence of the \Zt field term $h$, and when $V > 2t$ we have a $1/2$-parton Mott insulating state. Such state remains stable even for weak values of the \Zt electric field term $2h < V$. On the other hand, when the \Zt electric field term is dominant, the individual particles are confined and form the meson Luttinger liquid. The third regime of fluctuating dimers, the parton-plasma like regime, is concentrated around the $2h = V$ line when $h,V \gg t$.

\begin{figure*}[t!]
\centering
\epsfig{file=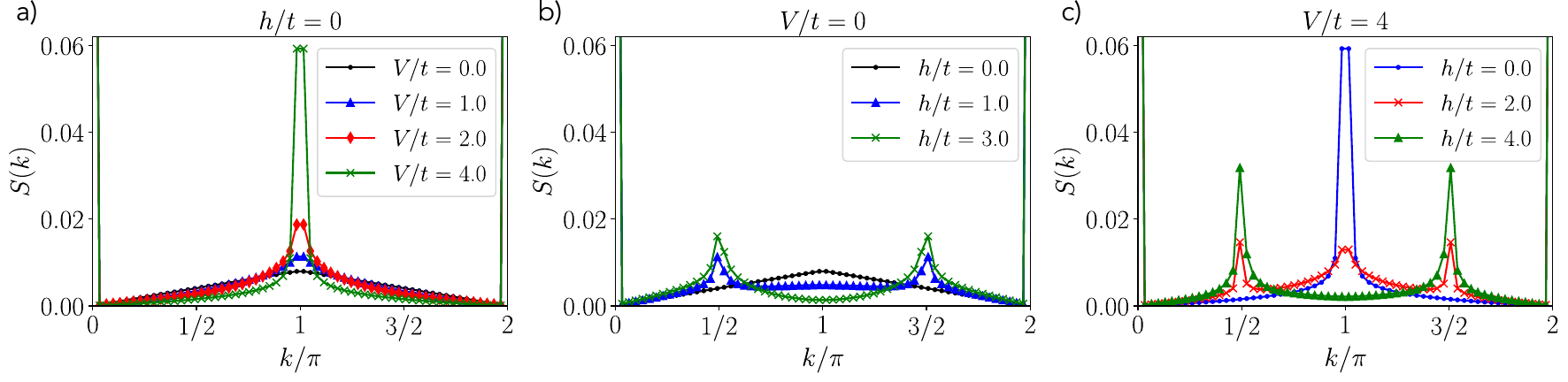, width=0.98\textwidth}
\caption{The static structure factor Eq.~\eqref{eq:StaticStructureFactor} for different values of parameters $h$ and $V$. a) In the absence of the electric field term $h = 0$, the peaks at $k = \pi$ rise significantly for $V > 2t$ signaling a Mott state. b) For zero value of the NN repulsion $V = 0$, the peaks at $k = \pi/2$ and $k = 3\pi/2$ become visible for the non-zero electric field term $h> 0$. c) At a finite value of the NN repulsion $V = 4t$, the peak at $k = \pi$ decreases with the increasing value of $h$, in contrast to the peaks at $k= \pi/2$ and $k = 3\pi/2$ which rise with the increasing strength of $h$. The height of both type of peaks are similar for $2h = V$.}
\label{fig:StructureFactorPeaks}
\end{figure*}

\section{Phase Diagram}

In this section we complete the phase diagram of the model. In the previous section we already uncovered some of the features in specific limits where analytical results are known and sketched the overal phase diagram in Fig.~\ref{fig:DiagramSketch}. Here we complement these results with numerical calculations. In particular we rely on the calculations of charge gap and static structure factor which allows us to distinguish different phases.

\subsection{Charge gap}

We considered the charge gap defined as
\begin{equation}
    \Delta(N, L) = \frac{E(N-2, L) + E(N+2, L) - 2 E(N, L) }{2},
    \label{eq:ChargeGap}
\end{equation}
in order to map out the Mott insulating state, where $E(N, L)$ is the ground state energy of a system with $N$ particles on $L$ lattice sites. Note that we consider $E(N \pm 2, L)$ states due to the Gauss law constraint where we make sure that every particle is part of a pair. Odd particle number would result in peculiar edge effects where a particle would stick to the edge.

The ground state energies are calculated using \textsc{SyTen} \cite{hubig:_syten_toolk, hubig17:_symmet_protec_tensor_network}, a DMRG toolkit \cite{White1992, Schollwoeck2011} (see also Appendix~\ref{NumericalDetails}). We fix the filling to one-half, $n = N / L = 1/2$, and calculate the ground state energies at different chain lengths in order to extrapolate the gap to the thermodynamic limit, $\Delta = \lim_{L \rightarrow \infty} \Delta(N, L)$ (see Appendix~\ref{ChargeGapDetails}). By scanning the values of parameters $h$ and $V$ we map out the phase diagram which is presented in Fig.\ref{fig:PhaseDiagram} a).

We obtain a Mott state, which is stabilised by the NN interactions for $V > 2t$ on the $h = 0$ line. This is in agreement with the analytical considerations in Sec.~\ref{AnalyticalLims_ZeroField}. Moreover we find an exponential opening of the gap which signals a BKT-like transition (see Appendix~\ref{ChargeGapDetails}). The Mott state is also observed for weak values of the electric field term relative to the NN interaction strength. It is clearly destabilized by stronger electric field term $h$. The gap disappears approximately at $2h = V$ which is again in agreement with our discussion on the fluctuating dimers in Sec.~\ref{AnalyticalFluctuating}.

\subsection{Static structure factor}
Motivated by the possible experimental implementation and to complement the charge gap results we also consider the static structure factor defined as \cite{Chanda2022}
\begin{equation}
    S(k) = \frac{1}{\mathcal{N}} \sum_{j,l} e^{-i(j-l)k} \left \langle \hat{n}_j \hat{n}_l \right \rangle.
    \label{eq:StaticStructureFactor}
\end{equation}
By using \textsc{SyTen} we calculated the structure factor by scanning through different values of $h$ and $V$ (see Appendix~\ref{StaticStructFctDetails}).
We observe two different types of peaks: at half-integer and integer multiples of $\pi$. These peaks depend on the values of our parameters $h$ and $V$, see Fig.~\ref{fig:PhaseDiagram} b) and c).

The peaks at $k = \pi$ show similar behaviour as the charge gap. They grow with the increasing value of $V$ and decrease by applying the \Zt electric field term, see Fig.~\ref{fig:StructureFactorPeaks} a) and b). This is in agreement with our predictions, namely that the peak at $k = \pi$ reflects the parton Mott state, where we expect that the configuration of the charges is an alternation of occupied and empty lattice sites (see
Appendix~\ref{StaticStructFctDetails}).

On the other hand, the structure factor peaks at ${k = \pi/2}$ show the exact opposite behaviour to the peaks at $k = \pi$, see Fig.~\ref{fig:StructureFactorPeaks} b). The height of the peaks rises with increasing value of the electric field term $h$, whereas no peaks at half-integer values of $\pi$ can be seen in the region where we expect a Mott state. Peaks at half-integer values of $\pi$ in combination with our earlier findings, where we observed no gap in that parameter regime, suggests that we observe the expected Luttinger liquid of confined dimers. Exponentially localized peaks at $k = \pi / 2 $ would generally mean that we have order with period of four lattice sites. However, the fact that we do not observe the charge gap in that parameter regime suggests that these are short-length correlations and the peaks are only algebraically localized. Hence, we do not observe a new Mott insulating state but a Luttinger liquid where the doubling of the period of Friedel oscillations in comparisons to $h = 0$ means that the constituents of the Luttinger liquid are confined mesons.

\section{Pre-formed Parton-Plasma}
\label{sec:parton_plasma}

As already discussed in Section \ref{AnalyticalLims} we can have energetically degenerate meson states at $2h = V$ in the absence of hopping.
These two meson configurations still dominate the physics when weak hopping is introduced as perturbation. Due to the degeneracy at $2h = V, t = 0$, the exact relation between the two competing interactions is important and results in a very different overall behaviour. Since we consider the commensurate filling of $n = 1/2$ we obtain a rather simple Mott state of alternating filled and empty lattice sites for $V \gg h$. This is reflected in the values of the charge gap as well as in the structure factor peak at $k = \pi$ discussed in the previous sections. On the other hand, when $h \gg V$ we obtain a Luttinger liquid of confined dimers where the NN interactions play a minor role. While these limits can be interpreted in terms of a Luttinger liquid of confined partons and a Mott state of individual partons respectively, the exact nature of the transition and the behaviour around the transition line is not that clear. This is due to the fact that the interplay of the \Zt electric field and the NN interaction results in the already mentioned frustrated state at $2h = V$ and the high level of doping. We witness a very interesting behaviour of individual $\mathbb{Z}_{2}-$charged particles which is hard to capture only in terms of the well known behaviour of the Mott state or the Luttinger liquid.

Now we investigate this regime more closely. The Mott transition occurs close to the line where $2h = V$ and $V \geq 2t$, as we already discussed in Sec.~\ref{AnalyticalFluctuating}. This is also reflected in the charge gap and in the static structure factor, see Fig.~\ref{fig:PhaseDiagram}. Near the transition line the confined mesons are able to fluctuate between length $l = 0$ and $l = 1$ configurations. The charge pairs are overall confined due to the non-zero value of the \Zt electric field term $h \neq 0$ \cite{Kebric2021, Borla2020PRL}. However, due to the high commensurate filling of $n = 1/2$ and the fact that fluctuations between the extended and tightly confined configuration are energetically almost equivalent, the particles effectively behave deconfined on short length scales in a sense specified below. Hence we label the region around the $2h = V$ line as a pre-formed parton-plasma region.

\begin{figure}[t]
\centering
\epsfig{file=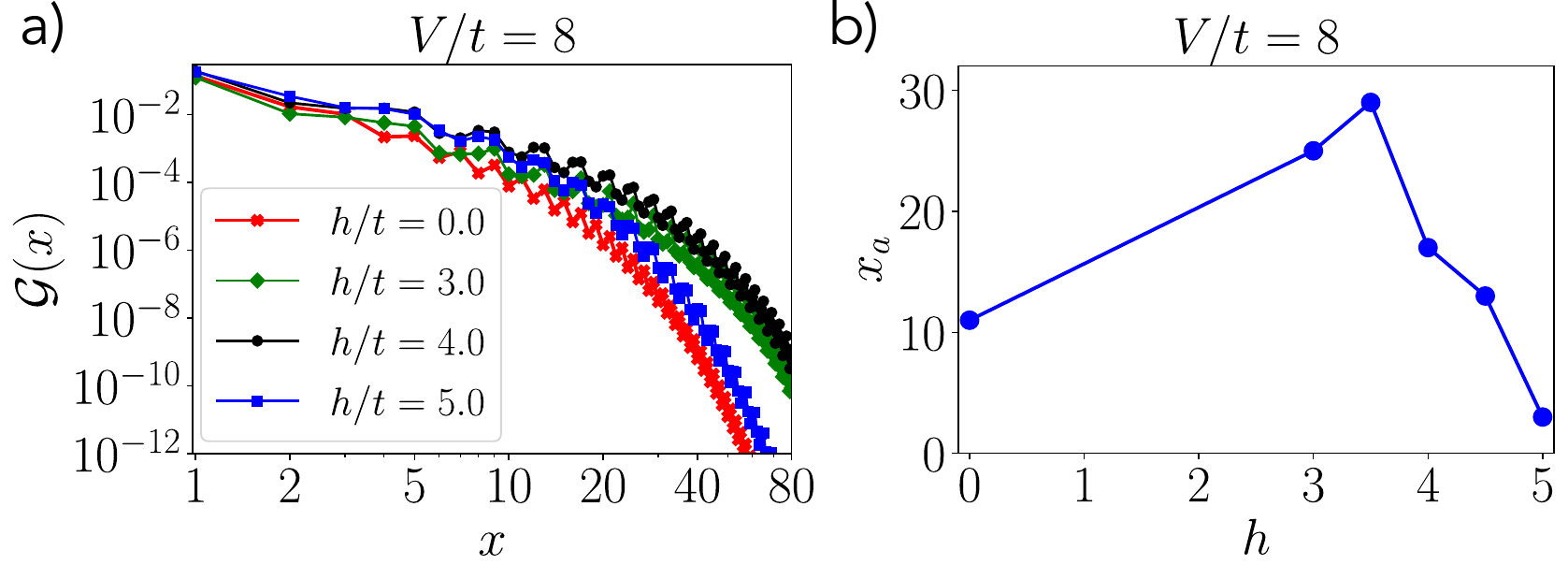, width=0.48\textwidth}
\caption{a) Green's function Eq.~\eqref{eq:Greens}, results calculated with \textsc{SyTen} in the Mott state (red x's), around the transition line at $2h = V = 8t$ (black dots), above the transition line $2h = 5t, V = 8t$ (green diamonds) and below the transition line $2h = 3t, V = 8t$ in the Mott state (blue squares). We observe strong exponential decay when $2h > V$ and a slower exponential decay when $2h < V$. This is contrasted by the case when $2h = V$ where the initial, slow, power-law decay eventually turns into a uniform exponential decay for longer distance. On the other hand a steady exponential decay is observed deep in the Mott state when $V = 8t$ and $h = 0$. b) Distance $x_a$ where the Green's function behaviour changes from algebraic to exponential decay as a function of $h$, for $V = 8t$.}
\label{fig:GreensPlasma}
\end{figure}

\subsection{Green's function}

In order to probe the confinement in the \Zt LGT model we consider the \Zt gauge-invariant parton Green's function \cite{Borla2020PRL, Kebric2021}
\begin{equation}
    \mathcal{G}(i-j) = \left \langle \ad_j \left( \prod_{ j \leq l < i} \tauZ_{ \langle l, l+1 \rangle} \right) \a_{i} \right \rangle.
    \label{eq:Greens}
\end{equation}
At long distances this Green's function decays exponentially for any non-zero value of the \Zt electric field term and algebraically for deconfined charges \cite{Borla2020PRL}.

On intermediate length-scales, in contrast, before the asymptotic long-distance behaviour is established, the parton Green's function can behave differently. In the strongly frustrated regime, $2h \approx V$, we can find power-law scaling on short-to-intermediate length scales resembling a parton plasma phase on length scales significantly exceeding the inter-particle spacing, see Fig.~\ref{fig:GreensPlasma}.

To analyze this behaviour more closely, we calculated the Green's function for different values of parameters $h$ and $V$ by using \textsc{SyTen}. The data was then fitted with a function containing a power-law decay, which reflects the deconfined behaviour on intermediate scales, and exponential decay which reflects confinement on long scales,
\begin{equation}
    f_f = A_f x^{-\alpha_f} e^{-\beta_f x}.
    \label{eq:FullFit}
\end{equation}
Here $x = |i-j|$ as defined in the Green's function, $\alpha_f$ denotes the strength of the power law decay and $\beta_f$ is a parameter which reflects the strength of the exponential decay. By considering the value of $\alpha_f$ and $\beta_f$ we quantify the character of the Green's function and distinguish between highly confined behaviour and deconfined behaviour.

\begin{figure*}[t]
\centering
\epsfig{file=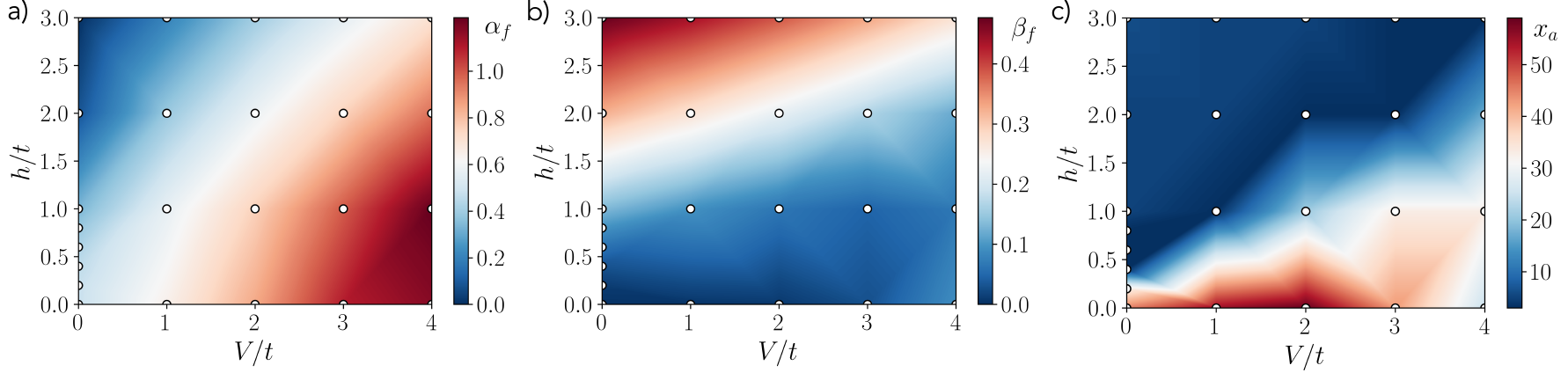, width=0.98\textwidth}
\caption{Heat maps of parameters $\alpha_f$ a) and $\beta_f$ b) obtained by fitting the DMRG data (white dots) of the parton Green's function with the fitting function for different values of parameters $h$ and $V$. c) Estimated value of the transition between the power-law decay and the exponential decay of the parton Green's function.}
\label{fig:HeatDiag_FullFits}
\end{figure*}

In order to perform the fit we took the logarithm of the data and the distance $x$ in order to capture the low values of the exponential decay better. As a result, we also had to transform the fitting equation Eq.~\eqref{eq:FullFit} accordingly (see Appendix~\ref{GreensFitsDetails}). The results of the fit can be seen in Fig.~\ref{fig:HeatDiag_FullFits}. For low values of $h$ the values of $\beta_{f}$ are very low and almost equal to zero. The values of $\alpha_f$ are substantially larger suggesting power-law decay consistent with the deconfined phase.
On the other hand, when $h$ becomes larger the values of $\beta_f$ start to rise and the value of $\alpha_f$ starts to decrease. The increase of $\beta_f$ appears to follow the $2h = V$ line which is in agreement with our previous numerical data (the charge gap and the structure factor) as well as our analytical arguments, that the particles become strongly confined into tight dimers. The fact that we can successfully extract two different parameters, which gradually change their values, already gives us the promising result that two different types of behaviour can be observed at the same time on different length scales. This can be clearly observed in Fig.~\ref{fig:GreensPlasma}, where we compare the Green's function results close to $2h = V$ line. The exponential decay prevails on the longer length scale whereas the initial decay appears to be power-law like (see Appendix~\ref{GreensFitsDetails} for more details).

Next, we determine the approximate position where the crossover from the algebraic to the exponential behaviour occurs by fitting the data only with a power-law function. In order to increase the precision of our fits we took the logarithm of our DMRG data which we then fitted with the appropriate logarithmic function that only captures the power-law decay
\begin{equation}
    f_c = A_c - \alpha_c \log(x).
    \label{eq:AlgLogFit}
\end{equation}
We fitted the above function to different number of data points $x$ for every parameter set $h, V$. In other words, we varied the number of data points $x_a$ considered in the fit ($1 \leq x \leq x_a$) from the minimum of $x_a = 3$ up to $x_a = 80$ (see Appendix~\ref{GreensFitsDetails}). We extracted the fit with the lowest sum of the absolute value of the covariance matrix elements which we determined as the best value of $x_a$. We then took this value as the position where the Green's function behaviour changes from power law decay to the exponential decay. The results deep in the Mott state at $V = 8t$ can be seen in Fig.~\ref{fig:GreensPlasma} b), where the algebraic behaviour is most prominent at around $h \approx 3.5 t$. This is slightly below the $2h = V = 8t$ line, however still consistent with our claim, since the exponential closure of the Mott gap occurs also slightly below the $2h = V$ line.

Results for general values of parameter $h$ and $V$ can be seen in 
Fig.~\ref{fig:HeatDiag_FullFits} c). 
The cutoff close to $x_a \approx 3$ means that the nature of the function is mostly exponential, signaling strong confinement on all length scales. This is indeed the case when $h \gg t$ as expected.
On the other hand, the cutoff at large values $ x_a \gtrsim 30$, signals a completely power-law behaviour, consistent with the deconfined phase. As can be seen in Fig.~\ref{fig:HeatDiag_FullFits} c) this is indeed the case for $h = 0$ and $V \leq 2t$ where the system is a deconfined Luttinger liquid which is again consistent with our previous claims.

The most interesting behaviour occurs when the cutoff value is on the intermediate length scale, i.e., when ${15 \lesssim x_a \lesssim 25}$. Such cutoff values would mean that the particles behave as if they were deconfined on short length scales $x\lesssim 10$. However on longer length scales, when $x\gtrsim 30$, the exponential decay prevails, which means that the particles are confined on longer length scales. Cutoff values in the range $15 \lesssim x_a \lesssim 25$ therefore signal pre-formed plasma-like behaviour. We extract such values of the cutoff close to the line where $2h = V$, which supports our previous analytical claims and numerical results, that we get plasma-like fluctuations on intermediate length scales.

We also briefly comment on the oscillation visible in the Green's function, see e.g. Fig.~\ref{fig:GreensPlasma}. Such oscillations are in fact predicted by the Luttinger liquid theory and depend on the effective filling of the chain \cite{Giamarchi2004}. We tried different fitting functions where we included the oscillations. The fits were generally successful deep in the confined Luttinger liquid regime and in the completely deconfined regime. However when the interplay between the electric field and NN interactions became stronger the quality of the fit became worse. Since we were not interested in the oscillations but in the general nature of the decay on different length scales we decided to exclude the oscillations in our fitting functions. As a result the quality of the fits became better and more reliable for a broader range of parameters. This is another hallmark of beyond-Luttinger liquid behaviour on short-to-intermediate length scales.

\begin{figure*}[ht]
\centering
\epsfig{file=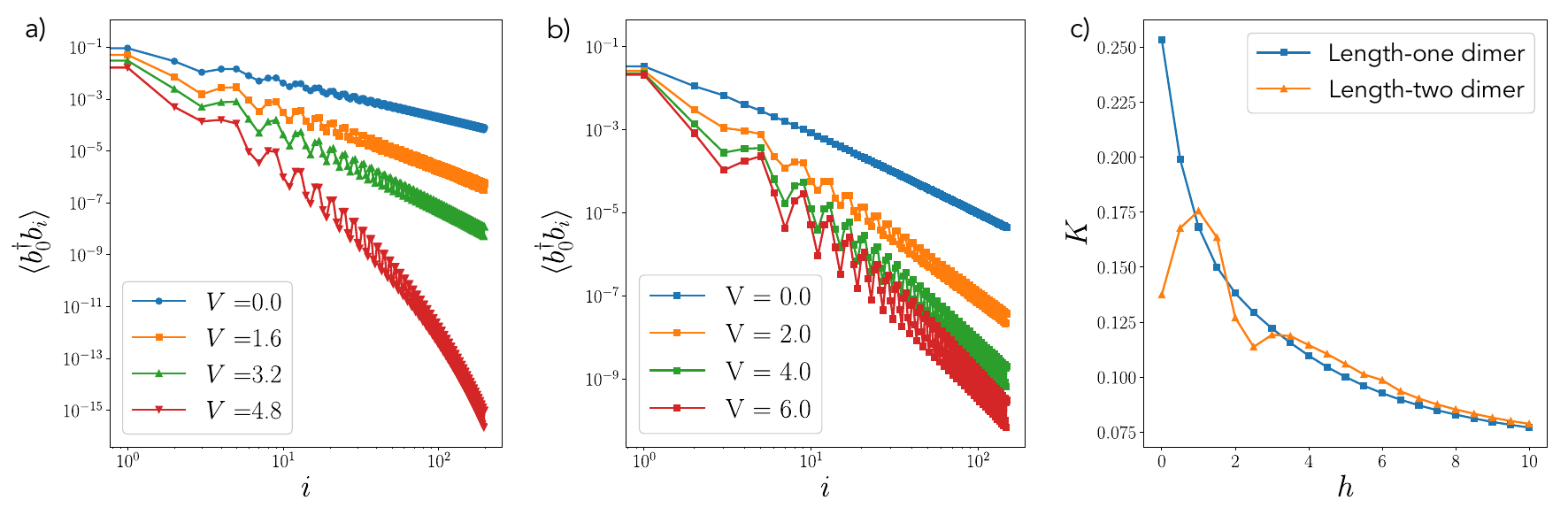, width=0.98\textwidth}
\caption{Pair-pair correlators at a fixed value of $h=2t$ (a) and on the $2h=V$ line (b), for several values of $V$. In the first case, the crossover from power-law to exponential decay indicates that the system becomes a Mott insulator for large enough $V$. In the second case, the correlators exhibit a power-law decay on the whole line, with an exponent $\alpha = \frac{1}{2K}$ that increases monotonically with $h$ as shown in (c). The same behaviour is also observed if we consider extended dimers of length two ($l = 1$). This hints at a gapless Luttinger liquid phase with strong repulsive interactions.
}
\label{fig:bb_corrs}
\end{figure*}

\subsection{Pair-pair correlators}

As discussed above, due to the confining nature of the $\mathbb{Z}_2$ electric field it is convenient in some regimes to consider the tightly bound mesons as the appropriate constituents. It is therefore useful to analyze the behavior of the gauge-invariant meson Green's function, which reads
\begin{equation}
    \langle \hat{b}^{\dagger}_i \hat{b}_j\rangle = \langle \hat{a}^{\dagger}_i \hat{\tau}^z_{i,i+1} \hat{a}^{\dagger}_{i+1} \hat{a}_j \hat{\tau}^z_{j,j+1} \hat{a}_{j+1}\rangle .
    \label{eq:bb_corr}
\end{equation}
We start by fixing $h$ to a fixed value, and evaluate the correlator \eqref{eq:bb_corr} numerically for different values of $V$, see Fig.~\ref{fig:bb_corrs}~a). Consistent with our expectations, it initially exhibits a power-law behavior but transitions to an exponential decay at a quantum critical point in the vicinity of $V=2h$. On one side of the transition we have a gapless Luttinger liquid, while on the other side the system becomes a Mott insulator and all the excitations are gapped. 

We then focus of the line $2h=V$ which, as explained in Section \ref{sec:parton_plasma}, is of particular interest due to the classical degeneracy between dimers of lengths one and two. In Fig.~\ref{fig:bb_corrs}~b) we observe that the correlators appear to exhibit a power law behavior on the whole line, with an exponent $\alpha$ that grows to large values as $h$ is increased, see Fig.~\ref{fig:bb_corrs}~c). 
We also calculated the same pair-pair correlations but for extended dimers, $l = 1$ (length-two dimers), where results are the same as for the tightly confined dimers $l = 0$ (length-one dimers) for stronger values of $h$.
If we stick to the Luttinger liquid paradigm, this corresponds to a very small Luttinger parameter $K=\frac{1}{2\alpha}$, which indicates strong repulsive interactions between the elementary constituents of the system \cite{Giamarchi2004}. Such result is surprising but can be possibly understood from the fact that at large $h$ the system is highly constrained, since the mesons cannot overlap. Similar systems, which are integrable, studied in \cite{Alcaraz_Bariev} exhibit power law decays with similarly large exponents.

\section{Conclusion and Outlook}
We mapped out the phase diagram of a half-filled one-dimensional \Zt LGT model, where dynamical U$(1)$ matter with repulsive NN interactions is coupled to a \Zt gauge field. We used analytical considerations and state of the art DMRG calculations and uncovered an interesting interplay between the confining non-local interaction, induced by the \Zt gauge field, and the NN repulsion.

The former exhibits a linear confining potential for parton pairs, which get effectively confined into mesons. They form a vast region of Luttinger liquid which remains stable even for small values of NN interaction. On the other hand we established a vast region where individual partons form a Mott insulating state induced by NN repulsive interaction alone. The \Zt electric field is able to destabilize the Mott insulating state due to the formation of tightly bound mesonic states. This happens once the field strength exceeds a threshold of $2h \approx V$ where two specific meson configurations are degenerate.

It is close to this line that the confined mesons are allowed to fluctuate under NN interaction, and give rise to a pre-formed parton-plasma region. Here, individual partons are still confined but the mesons are allowed to fluctuate to a dimer where there is an empty lattice site between two partons. This is allowed due to the NN interaction which compensates for the linear confining potential of the \Zt electric field term. Combining this effect with high doping at one-half filling we obtain an apparent absence of confinement on short-to-intermediate length scales. This causes plasma like behaviour of individual partons. However, on longer length-scales the partons are still confined. This behaviour is observed in the parton Green's function and in the pair-pair correlation function.

The remaining question which is yet to be answered is the precise nature of such transition. For this, our effective squeezed model might be utilized. An interesting extension would also be to reconsider the dynamical charges as spinful particles, or consider our model in higher dimensions, and at non-zero temperature.

 Moreover it would be desirable to compare our results to experiments. In particular the Rydberg dressing scheme where the model maps exactly to a $t-J_z$ model \cite{Kebric2021} is a promising approach which could be implemented in a cold atom experiment in addition to the already mentioned Floquet scheme \cite{Barbiero2019, Goerg2019, Schweizer2019}, superconducting qubits \cite{Zohar2017, HomeierPRB2021} and recent Rydberg tweezer array proposals \cite{Halimeh2021DisorderFree, Halimeh2021LPG}.

\section*{Data availability statement}

The data that support the findings of this study are available upon reasonable request from the authors.

\begin{acknowledgments}
We thank T. Zache, J. Halimeh, A. Bohrdt, L. Homeier, and F. Palm for fruitful discussions.

This research was funded by the Deutsche Forschungsgemeinschaft (DFG, German Research Foundation) under Germany's Excellence Strategy -- EXC-2111 -- 390814868 and via Research Unit FOR 2414 under project number 277974659, and received funding from the European Research Council (ERC) under the European Union’s Horizon 2020 research and innovation programm (Grant Agreement no 948141) — ERC Starting Grant SimUcQuam.
UB is funded by the Deutsche Forschungsgemeinschaft (DFG, German Research Foundation) under Emmy Noether Programme grant no. MO 3013/1-1 and under Germany’s Excellence Strategy - EXC-2111 - 390814868.
SM is supported by Vetenskapsradet (grant number 2021-03685).

\end{acknowledgments}


\appendix

\section{Mapping to the XXZ model}\label{SuppXXZmodel}

The hard-core bosonic model \eqref{eq:SLFermionicHamiltonian} can be mapped to the XXZ spin-$1/2$ chain by a simple mapping where the bosonic creation and annihilation operators are replaced with spin raising and lowering operators ${ \hat{b}^{\dagger}_j \rightarrow \hat{S}^{+}_{j} }, {\hat{b}_{j}  \rightarrow \hat{S}^{-}_{j} }, {\hat{b}_j^\dagger \hat{b}_j - 1/2 \rightarrow \hat{S}^z_j }$ \cite{Giamarchi2004}.
Such mapping gives us the XXZ model up to constant factors \cite{Giamarchi2004}
\begin{equation}
    \H_{\rm XXZ} = \sum_j \left [ J_{\rm xy}  \left( \hat{S}^x_{j+1} \hat{S}^x_j + \hat{S}^y_{j+1} \hat{S}^y_j \right) + J_{\rm z} \hat{S}^z_{j+1} \hat{S}^z_j \right ].
\end{equation}
The parameters in the XXZ model are directly related to the parameters in the original \Zt LGT model as $2t = J_{\rm xy}$ and $V = J_{\rm z}$ \cite{Giamarchi2004}. Due to the mapping, ${\hat{b}_j^\dagger \hat{b}_j - 1/2 \rightarrow \hat{S}^z_j }$, the magnetization in the $z-$direction of the XXZ model is directly related to the filling in the bosonic model \eqref{eq:SLFermionicHamiltonian}. Hence a magnetic field term in the $z$-direction in the XXZ model is related to the chemical potential term in the bosonic model.

The diagram for generic values of the parameters ($J_{\rm xy}$ and  $J_{\rm z}$) and filling can be calculated via the Bethe ansatz \cite{Giamarchi2004, Sutherland2004, Haldane1980, Luther1975}. For zero magnetization it can be calculated analytically. There the model is known to have an antiferromagnetic transition point at $J_{\rm z} = J_{\rm xy}$, meaning that the system is in the gapped antiferromagnetic phase for $J_{\rm z} > J_{\rm xy}$ and forms a Luttinger liquid for $J_{\rm z} < J_{\rm xy}$ \cite{Giamarchi2004}. 
The half-filling of the original model~\eqref{eq:SLFermionicHamiltonian} corresponds to zero magnetization in the XXZ spin-1/2 chain since there are equal number of up and down spins which corresponds to the equal number of occupied and empty sites. Hence we expect a transition to the gapped Mott state in the bosonic model \eqref{eq:SLFermionicHamiltonian} at $V = 2t$, since $2t = J_{\rm xy}$ and $V = J_{\rm z}$. There is no transition when the system is away from zero magnetization for any value of $J_{\rm xy}$ and $J_{\rm z}$ \cite{Giamarchi2004}. Hence, considering the same parameter regime, but slightly away from half-filling, would result in the system remaining a Luttinger liquid and there would be no transition to a Mott state.

\section{Details on the fluctuating dimers} \label{FluctDimers}

In this section we present more details on the derivation of the effective model \eqref{eq:EffectiveSpinfulwithField} via the squeeze of the two different types of dimers into spinful particles. For simplicity we first consider $2h = V$ and $h, V \gg t$. The first condition means that there is energetically no difference between tightly confined and extended dimer. In the end we will comment what happens when we relax the condition to $2h \approx V$, but still keep $h, V \gg t$.

\emph{Hopping term.--}
We first focus on the hopping of the dimers. We consider an extended dimer depicted in Fig.~\ref{fig:DimerSqueeze}~d). Due to our imposed restriction, $l<2$, we realize that the left particle of such dimer can hop to the right and that a right particle can hop to the left. In other words: such dimer can only become shorter, since the only allowed hopping is for one of the particles in the dimer to hop towards the other particle in the same dimer. In the first case we obtain the configuration depicted in Fig.~\ref{fig:DimerSqueeze} f) and in the later case we get the configuration sketched in Fig.~\ref{fig:DimerSqueeze} e). 
We observe how is such hopping reflected in the squeezed spin language, sketched next to the original \Zt LGT configurations, in Fig.\ref{fig:DimerSqueeze} d)-f).
Contraction of the extended dimer to the right means that a pseudospin-up flips to a pseudospin-down on the same site and a contraction to the left means that a pseudospin-up hops to the right \emph{and} flips to a pseudospin-down. Opposite processes are of course equivalent. On the other hand, any other hopping is forbidden since it does not correspond to any allowed physical configuration, where the dimer remains $l<2$. Summarizing all of the above allows us to write the hopping term in the spinful language as
\begin{equation}
 	\hat{H}_{s}^{t} = \hat{P}_{d} \left[ - t \sum_{j} \left ( \hat{c}^{\dagger}_{\uparrow, j}  \hat{c}_{\downarrow, j+1} + \hat{c}^{\dagger}_{\uparrow, j}  \hat{c}_{\downarrow, j} + h.c. \right )\right]
 	\hat{P}_{d},
	\label{eq:EffectiveDimerKinetic}
\end{equation}
where the operator $\hat{P}_d$ projects on the subspace where there are no double occupancy on the same lattice site.

\emph{Potential terms.--}
Since we consider $V \gg t$ we exclude all nearest neighbour configurations of particles which are not part of the same dimer. This doesn't include particles in the same dimer, where we have the competition between the \Zt electric string tension and the NN repulsion. The energy difference between the extended dimer and the tightly confined dimer equals to $\delta = 2h - V$ as was already defined in the main text. Since we will generally consider $2h \approx V$, this means that $\delta \ll 2h, V$ and we therefore allow NN configurations in the \textit{same} dimer.

Some of the NN configurations are already restricted by our construction. For example two neighboring pseudospins-down or pseudospin-down followed by any other particle has an empty site between charges per our construction. Hence only configurations for which we still need to impose high energy penalty are configurations of a pseudospin-up followed by any other pseudospin
\begin{equation}
 	\hat{H}_s^V = \hat{P}_{d} \left[ V \sum_{j} \hat{n}_{\uparrow, j} \hat{n}_{j+1} \right] \hat{P}_{d}.
	\label{eq:EffectiveDimerPotential}
\end{equation}
As already stated before, such states are energetically heavily penalised and we can project them out by promoting the $\hat{P}_{d} \rightarrow \hat{P}$ where $\hat{P}$ projects to the subspace where also the configurations contributing to terms in Eq.~\eqref{eq:EffectiveDimerPotential} are projected out in addition to the projection to the subspace where there are no double occupancy on the same lattice site.

\begin{figure}[t]
\centering
\epsfig{file=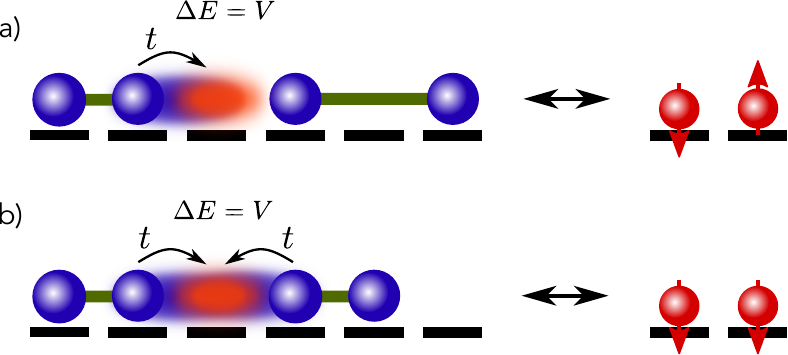, width=0.48\textwidth}
\caption{Two states where dimers are allowed to occupy nearest neighbor sites. However any hopping of the particle in the left dimer to the right is forbidden in case a) and either hopping of a particle towards the neighbouring dimers is forbidden in case b). Both of these states contribute to additional terms, Eq.~\eqref{eq:secondOrderCorrections}.}
\label{fig:SecondOrderPerturbNNint}
\end{figure}

\emph{Perturbative terms.--}
Finally, we consider the perturbative corrections due to restricted hopping, see Fig.~\ref{fig:SecondOrderPerturbNNint}. Both of the depicted configurations in Fig.~\ref{fig:SecondOrderPerturbNNint} are allowed and are not penalized by any NN interaction. However such configurations have restricted hopping. As a result we have to add additional NN interactions which are second order perturbation corrections in hopping
\begin{equation}
    \H_s^{2nd} = \frac{t^2}{2V} \sum_{j} \hat{n}_{\downarrow, j} \hat{n}_{\uparrow, j+1} + \frac{t^2}{V} \sum_{j} \hat{n}_{\downarrow, j} \hat{n}_{\downarrow, j+1}
    \label{eq:secondOrderCorrections}.
\end{equation}

\begin{figure*}[th]
\centering
\epsfig{file=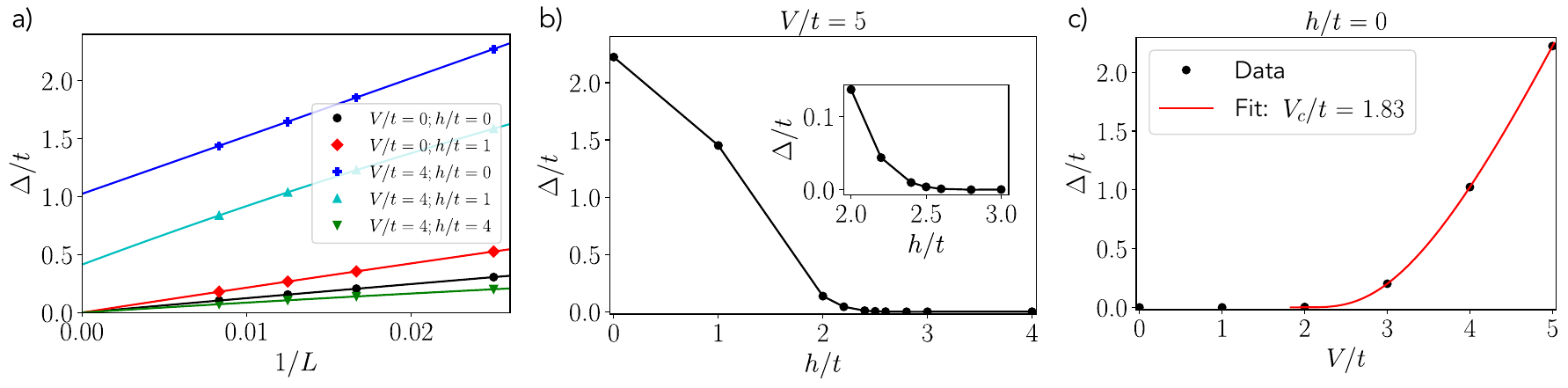, width=0.98\textwidth}
\caption{a) Typical plots of the finite size scaling of the charge gap with a quadratic function. b) Charge gap opening as a function of $h$ for the constant value of the NN interactions $V = 5t$. The inset in b) shows the gap closing for $h > 2.5t$. c) BKT opening of the charge gap \eqref{eq:ChargeGap} in thermodynamic limit as a function of $V$ for $h = 0$.}
\label{fig:GapDetailsSupp}
\end{figure*}

Combining all terms together gives us the effective model for fluctuating dimers
\begin{multline}
 	\hat{H}_s = \hat{P} \left[ - t \sum_{j} \left ( \hat{c}^{\dagger}_{\uparrow, j}  \hat{c}_{\downarrow, j+1} + \hat{c}^{\dagger}_{\uparrow, j}  \hat{c}_{\downarrow, j} + h.c. \right ) \right. \\
 	\left. +\frac{t^2}{2V} \sum_{j} \hat{n}_{\downarrow, j} \hat{n}_{\uparrow, j+1} + \frac{t^2}{V} \sum_{j} \hat{n}_{\downarrow, j} \hat{n}_{\downarrow, j+1} \right] \hat{P}.
	\label{eq:EffectiveSpinfulDegDimers}
\end{multline}
We once again note that we considered $V = 2h$ meaning that the two dimers are energetically equivalent. We can relax such condition to $2h \approx V$ while still being in the limit where $h, V \gg t$. In such case the equation acquires one extra term, which comes directly from the energy difference between the two types of dimers
\begin{multline}
 	\hat{H}_s = \hat{P} \left[ - t \sum_{j} \left ( \hat{c}^{\dagger}_{\uparrow, j}  \hat{c}_{\downarrow, j+1} + \hat{c}^{\dagger}_{\uparrow, j}  \hat{c}_{\downarrow, j} + h.c. \right ) + \right. \\
 	\left. \frac{t^2}{2(\delta + V)} \sum_{j} \hat{n}_{\downarrow, j} \hat{n}_{\uparrow, j+1}
 	+ \frac{t^2}{\delta + V} \sum_{j} \hat{n}_{\downarrow, j} \hat{n}_{\downarrow, j+1} \right. \\
 	\left. + \delta \sum_{j} n_{\uparrow, j} \right] \hat{P}.
	\label{eqEffectiveSpinfulwithFieldSupplement}
\end{multline}

For $\delta > 0$ we expect a spin imbalance meaning that most of the dimers are short. In addition we also have a second order NN interaction. This interaction is always smaller than the hopping amplitude $\frac{t^2}{\delta + V} = \frac{t^2}{2h} \ll t$ since $h,V \gg t$. This would suggest that the ground state could not enter a Mott state of alternating empty sites and down spins, which in the \Zt language means a state where two empty sites are followed by two occupied sites. However we note that also the hopping amplitude would have to be corrected due to the occupational imbalance of the up and down particles. On the other hand increase of the \Zt electric field $h \rightarrow \infty$, also diminishes the effect of NN interactions and gives us the Luttinger liquid of confined dimers.

The dimers will energetically favour the extended state when considering the opposite case $\delta < 0$. We are thus getting close to a simple Mott state of alternating occupied and empty lattice sites. This is consistent with the opening of the charge gap in the thermodynamic limit and the peaks in the structure factor at $k = \pi$, see Fig.~\ref{fig:PhaseDiagram} in the main text.

\section{Details on the numerical calculations}\label{NumericalDetails}

Numerical calculations of the charge gap, static structure factor, and the \Zt parton Green's function were performed using \textsc{SyTen} \cite{hubig:_syten_toolk, hubig17:_symmet_protec_tensor_network}, which is a DMRG toolkit. More specifically we used the \textit{pyten} module of the \textsc{SyTen} \cite{hubig:_syten_toolk}.
In order to carry out the calculations we map the original model Eq.~\eqref{eq:LGT_Model} to a spin$-1/2$ model by using the Gauss law constraint \cite{Kebric2021, Borla2020PRL, Lange2022}. More precisely, we use the constraint $\hat{G}_j = +1$ in order to write $\hat{n}_j = \frac{1}{2} \left ( 1 - 4 \hat{S}^{x}_{j} \hat{S}^{x}_{j+1} \right )$, which allows us to rewrite the original model entirely with spin-$1/2$ variables \cite{Kebric2021, Borla2020PRL, Lange2022}
\begin{multline}
    \H_s = t \sum_{j = 2}^{\tilde{L}-1} \left( 4 \hat{S}_{j-1}^{x}\hat{S}_{j+1}^{x} - 1 \right) \hat{S}_{j}^{z}
    - 2 h \sum_{j=1}^{\tilde{L}} \hat{S}_{j}^{x}\\
    + V \sum_{j=2}^{\tilde{L}-1}
    \frac{1}{2} \left( 1 - 4 \hat{S}_{j-1}^{x}\hat{S}_{j}^{x} \right)
    \frac{1}{2} \left( 1 - 4 \hat{S}_{j}^{x}\hat{S}_{j+1}^{x} \right) \\
    + 2 \mu \sum_{j=1}^{\tilde{L}-1} \hat{S}_{j}^{x}\hat{S}_{j+1}^{x},
\end{multline}
Here $\tilde{L}$ is the number of \Zt gauge and electric fields - link variables. Such mapping was already implemented in Refs.~\cite{Kebric2021, Borla2020PRL, Lange2022}. Similar model was also considered to study quantum scared states \cite{Iadecola_2020, Halimeh2022QuantumScars, Aramthottil2022}. We added the chemical potential term in order to control the filling. We consider open boundary conditions and therefore start and end our lattice with a link variable. Hence, we simulate $\tilde{L} = L + 1$ spins. The same trick was used also for the calculations of the correlation function; i.e. the Green's function and the density correlations for the static structure factor. Convergence was determined by comparing energy difference between two consecutive sweeps, which had to be lower than $\sim 2 \cdot 10^{-15}$.  In addition we also checked that the variance of the Hamiltonian was low in order to confirm that the state is an eigenstate.

The pair-pair correlations in Fig.~\ref{fig:bb_corrs} were calculated using the iDMRG algorithm, implemented through the tensor network package TeNPy \cite{tenpy}. In this case we simulate directly the model \eqref{eq:LGT_Model}, enforcing the Gauss law energetically. By employing the charge conservation library of TenPy  the system can be kept exactly at half-filling with no need to tune the chemical potential. This is crucial for the study of gapless phases, especially in the large $h$ limit

\section{Details on the charge gap calculations} \label{ChargeGapDetails}

\emph{Finite size scaling.--}
In order to extract the charge gap in the thermodynamic limit we extrapolate the charge gap values to infinite system lengths, ${\Delta= \lim_{L \rightarrow \infty}=\Delta(N, L)}$. This was done by plotting the charge gap, Eq.~\ref{eq:ChargeGap}, as a function of the inverse system length, $1/L$, and fitting it with a quadratic function. Some typical examples of the extrapolation for different parameters together with the fits can be seen in Fig.~\ref{fig:GapDetailsSupp}~a). The value of the quadratic function at $1/L = 0$ was taken as the charge gap in the thermodynamic limit. Typically, we took the system lengths of $L = 40, 60, 80, 120$. For higher values of $h$ lower system sizes had to be taken in order to ensure convergence of the DMRG. For $h = 4t$ convergence became challenging already for $L = 60$. This is because large electric field value $h$ also means that the value of the corresponding chemical potential $\mu$ had to be increased in order to remain at half-filling.

\begin{figure*}[th]
\centering
\epsfig{file=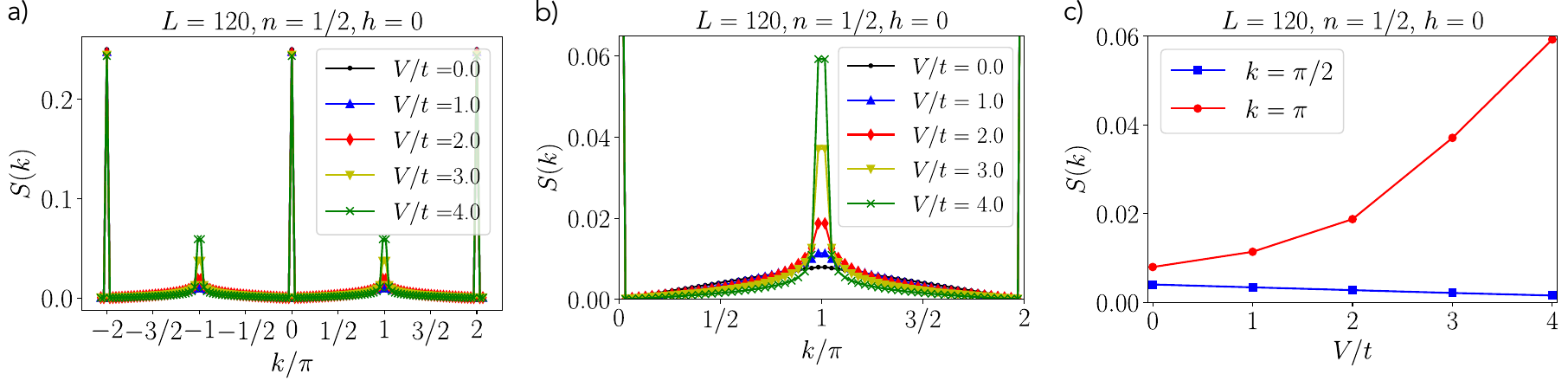, width=0.98\textwidth}
\caption{a) Structure factor for different values of $V$ for zero field $h = 0$ for the Hamiltonian \eqref{eq:LGT_Model}. b) A prominent peak is obtained for $V \geq 2t$ at $k = \pi$ consistent with the Mott state expected in such regime. c) Details on amplitude of the peaks at $k = \pi/2$ and $k = \pi$ c) as a function of $V$ for $h = 0$.}
\label{fig:ZeroFieldStructureDetailed}
\end{figure*}

\emph{Gap closing at $2h = V$.--}
As can be seen in Fig.~\ref{fig:PhaseDiagram} a) the charge gap generally closes when $2h > V$. This is in agreement with our analytical considerations in Sec.\ref{AnalyticalLims}. In Fig.~\ref{fig:GapDetailsSupp}~b) we show details on the gap closing as a function of the \Zt electric field term $h$, when the NN interactions are fixed at a substantial value $V = 5t$. The inset in Fig.~\ref{fig:GapDetailsSupp}~b) shows that the gap opens for $h > 2.5 t$ supporting our claim that the Mott state is destabilized for $2h > V$.

\emph{BKT gap opening at $h = 0$.--}
When the \Zt electric field term is zero $h = 0$, the LGT model \eqref{eq:LGT_Model} maps to a hard-core bosonic model which undergoes a Mott transition for $V > 2t$, see the Main text and the Supplementary material Sec.~\ref{SuppXXZmodel} for details. We verified our numerical results by fitting the gap opening as a function of $V$ with the BKT exponential function \cite{Giamarchi2004, Sutherland2004}
\begin{equation}
    \Delta_{\rm BKT} = C_1 e^{\frac{-C_2}{\sqrt{V-V_c}}},
    \label{eq:BKTGap}
\end{equation}
where $V_c$ is the transition point. We had to fix the value of $V_c$ by hand and fitted the function Eq.~\eqref{eq:BKTGap} where $C_1$ and $C_2$ were free parameters. The result of the best fit can be seen in Fig.~\ref{fig:GapDetailsSupp} c). The value of the transition point for which we obtained the best fit was $V_c/t = 1.83 \pm 0.03$. This is less than $9\%$ off from the analytically expected value at $V_c = 2t$, which means that our numerical calculations are consistent with analytical results. The reason for a slightly lower value for $V_c $ comes from the very nature of the fitting function which we used. Since some small errors  accumulated during the finite size scaling it was very hard to determine when exactly does the gap open since we were dealing with very small values, which had to be fitted with an exponential function in Eq.~\eqref{eq:BKTGap}.

\section{Details on the static structure factor} \label{StaticStructFctDetails}

Typically we calculated the ground state of a chain with $L = 120$ sites. In order to avoid edge effects we only consider sites in the middle, $ 30 < j, l \leq 91$. Hence $k$ is the discretized momentum defined as $k = \frac{s}{\tilde{L}} \pi$, where $s \in \mathbb{Z}$ and $\tilde{L}$ is the number of considered data points points ($\tilde{L} = 61$ for $L = 120$ since we discarded first $30$ and last $29$ data points. As a result the normalization factor in the main text takes the value $\mathcal{N} = \tilde{L}^{2}$.

In our numerical results of the finite DMRG we always observe strong peaks at the origin which also repeat with the period $2\pi$, see Fig.~\ref{fig:ZeroFieldStructureDetailed} a), where the structure factor for extended values of $k$ is plotted. We associate such peaks to the fact that we consider the non-connected density-density correlations when doing the Fourier transformation for the static structure factor.

In addition to that, we also observe peaks at $k = \frac{\pi}{2}$, $k = \pi$, and $ k = \frac{3\pi}{2}$ for different set of parameters $h$ and $V$. We estimate the height of these peaks and compare it to the charge gap diagram in Fig.~\ref{fig:PhaseDiagram} a). We note that the discretization of $k$ is connected to the number of data points in our structure factor. Hence our values of $k$ are not exactly $k = \frac{\pi}{2}$, $k = \pi$ and $k = \frac{3\pi}{2}$. However in the chains with length $L = 120$  the number of considered data points was $\tilde{L} = 61$ and the values of $k$ for which we have the available value of the structure factor are rather close to the exact value. Although a systematic error is made in such procedure we are interested in the qualitative change of the height of the peaks. Such qualitative behavior is thus not spoiled since we always take the same $k \approx \frac{\pi}{2}, \pi, \frac{3\pi}{2}$ for different parameters. Hence by extracting the values of the structure factor at such $k$, we can plot a phase diagram for different parameters of $h$ and $V$. The results can be seen in Fig.~\ref{fig:PhaseDiagram} b) and c) in the main text.

In the following we consider the numerical results for different parameter regimes separately.

\emph{Zero electric field term.--}
In the case when the \Zt electric field term is set to zero $h = 0$ and $V \geq 0$, we observe strong peaks at $k = \pi$, and no peaks at $k = \frac{\pi}{2}, \frac{3\pi}{2}$, see Fig.\ref{fig:PhaseDiagram} b) and c). Detailed plots of the static structure factor as a function of $k$ for different values of $V$ are also presented in Fig.~\ref{fig:ZeroFieldStructureDetailed}. The amplitude of the peak rises significantly for $V/t > 2$, which is presented in  Fig.~\ref{fig:ZeroFieldStructureDetailed} c). Such behaviour points to a long range staggered charge configuration where every other lattice site is occupied. This agrees perfectly with the Mott insulating state, which was already established by calculating the charge gap (see Fig.~\ref{fig:PhaseDiagram} a)) and our analytical considerations in Sec.~\ref{AnalyticalLims_ZeroField}. 

\begin{figure}[th]
\centering
\epsfig{file=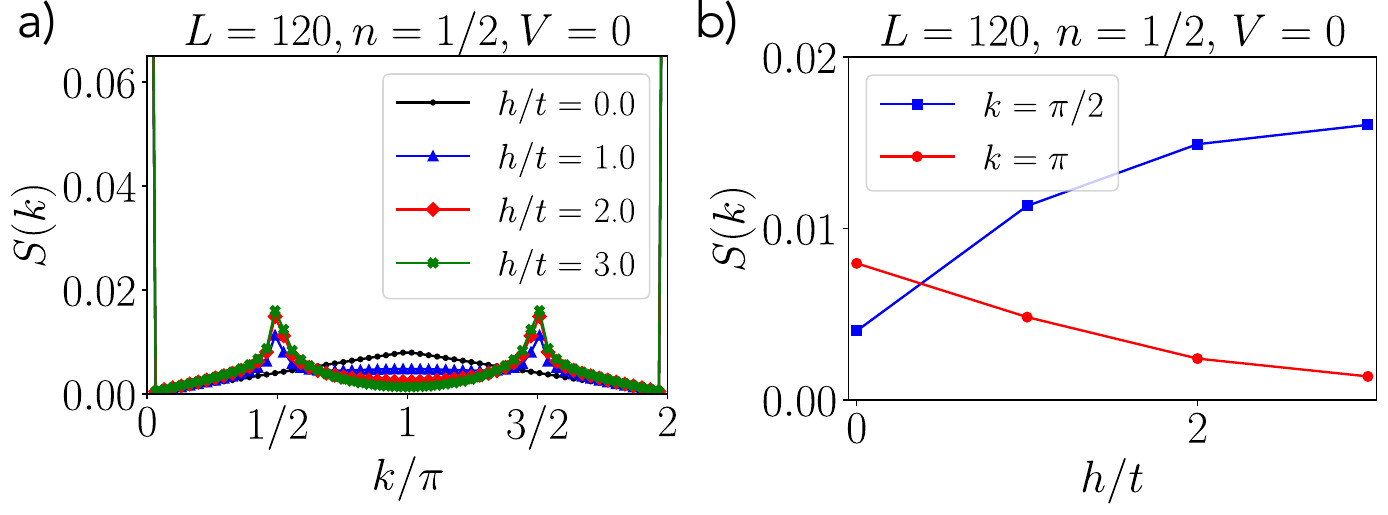, width=0.48\textwidth}
\caption{a) Structure factor for different values of $h$  and constant $V = 0$ for the Hamiltonian \eqref{eq:LGT_Model}. A prominent peaks are obtained for $h > 0$, $V = 0$ at $k = \frac{\pi}{2}$ and $k = \frac{3\pi}{2}$. b) The height of the peak as a function of $h$ at $V = 0$ at $k=\pi/2$ and $k = \pi$ respectively.}
\label{fig:ZeroNNStructure}
\end{figure}

\begin{figure}[th]
\centering
\epsfig{file=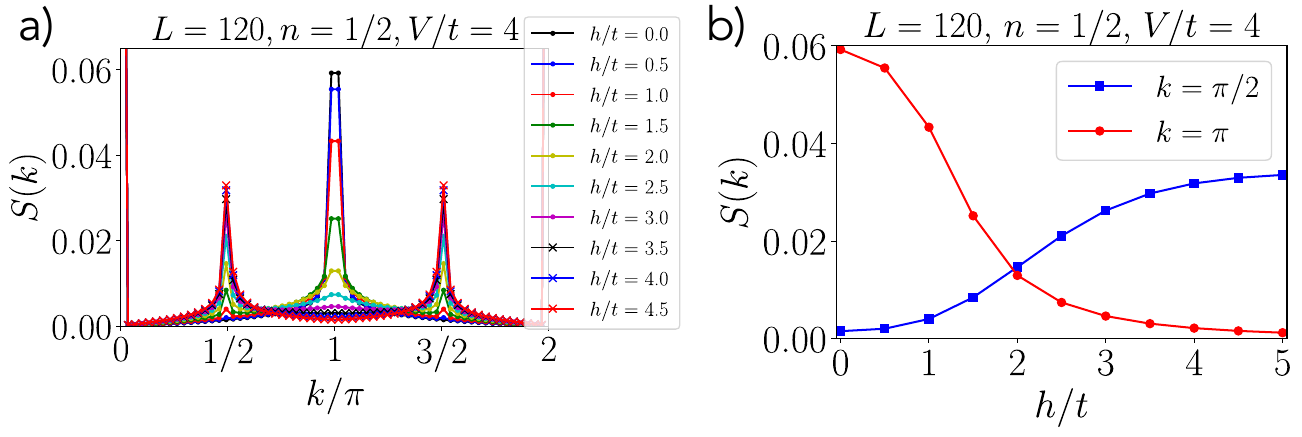, width=0.48\textwidth}
\caption{a) Structure factor for different values of $h$ and constant NN repulsion $V = 4t$ for the Hamiltonian \eqref{eq:LGT_Model}. Peaks at $k = \pi/2$ and $k = 3 \pi /2$ appear when we include the \Zt electric field term. High peaks at $k = \pi$ start to decrease with the increasing value of $h$ and the peaks at half-integer $k$ increase with the increasing value of $h$. Both peak types appear to be of the same magnitude as the integer peaks at $2h = V$. b) Height of the peak at different values of $k$ as a function of $h$ at $V = 4t$.}
\label{fig:NNand_h_Structure}
\end{figure}

\begin{figure*}[th]
\centering
\epsfig{file=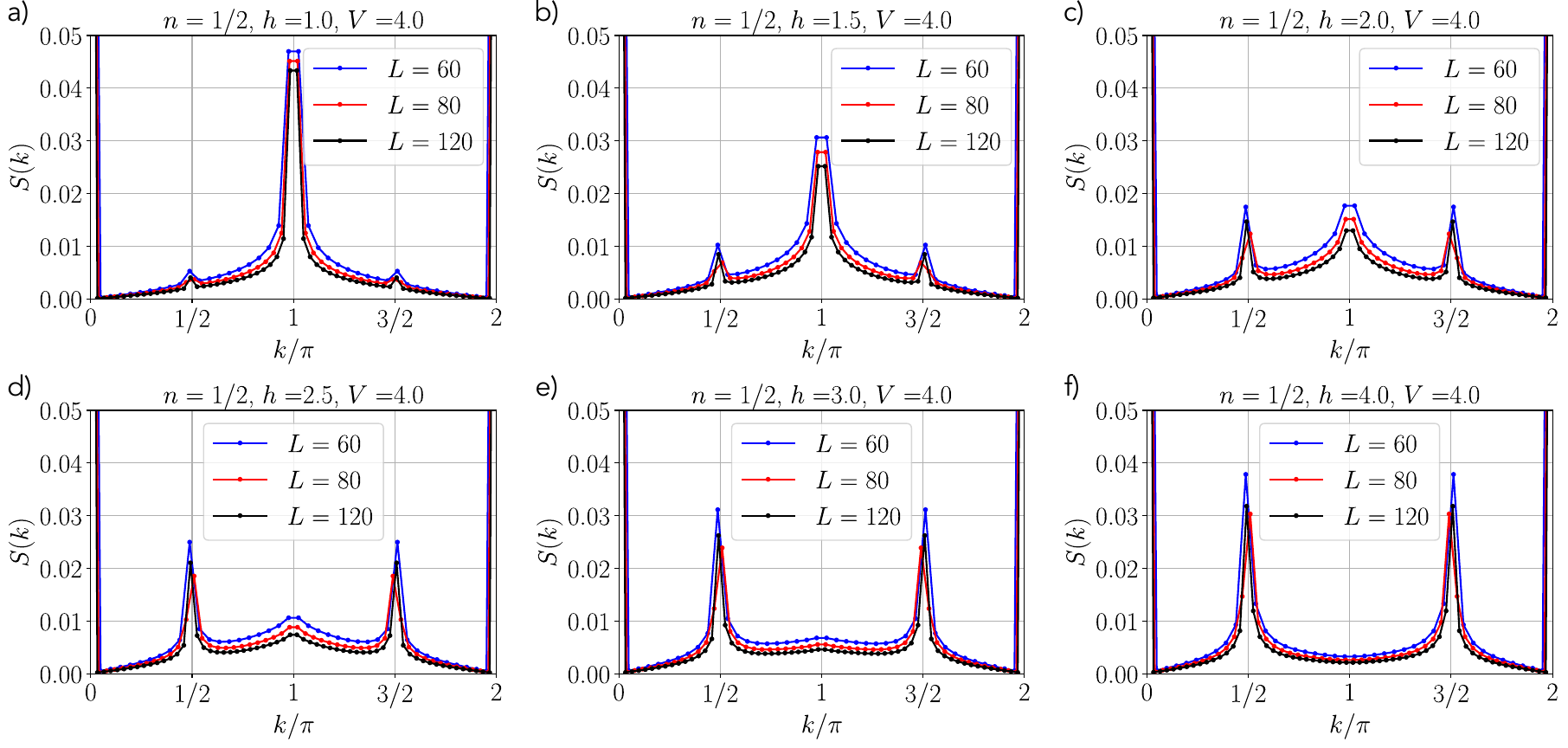, width=0.98\textwidth}
\caption{Structure factor for different chain lengths $L$ and different parameters $h$ and $V$.}
\label{fig:FiniteSizeEffectStructure}
\end{figure*}

\emph{Tightly confined dimers.--}
Next, we consider the structure factor where the NN interactions are set to zero and we vary the \Zt electric field term strength. The results can be seen in Fig.~\ref{fig:ZeroNNStructure}. In such regime we observe somehow modest peaks at $k = \pi / 2$ and no peaks at $k = \pi$ which would signal Mott state as in the case when $V > 2t$. The observed peaks, however, indicate that the periodicity of the charge-charge correlations doubled. This is in agreement to the observations that the Friedel oscillations double for any nonzero value of the \Zt electric field term \cite{Borla2020PRL}. However the peaks are not very high and we expect them to become even less prominent in the thermodynamic limit for $h \gg t$, where we expect Luttinger liquid of confined dimers and thus no true long-range correlations.

\emph{Fluctuating dimers.--}
We also consider the structure factor when we include both the NN interaction and the \Zt electric field term. In particular we set the NN interaction to $V = 4t $ deep in in the Mott state and varied $h$. As can be seen in Fig.~\ref{fig:NNand_h_Structure}~a) substantially high peaks at $k = \pi$ start to decrease with the increasing value of $h$. On the other hand peaks at $k = \pi/2$ and $k = 2\pi/3$ appear already for very small values of $h$ and their amplitude increases with increasing value of $h$. The change of the amplitude of the peaks is best seen in Fig.~\ref{fig:NNand_h_Structure}. We see that the amplitudes of both peaks are approximately the same at $2h = V$. Such crossover in peak behaviour is once again in agreement with the closing of the Mott gap at $2h = V$. The coexistence of the peaks signals a very interesting regime.

\emph{Finite size effects in $S(k)$.--}
Finally, we also check the importance of the finite size effects in our numerical calculation. We calculated the structure factor for smaller system sizes and compared the results. The results for $L = 60$, $L = 80$ and $L = 120$ can be seen in Fig.~\ref{fig:FiniteSizeEffectStructure}. Generally, the height of the peaks are slightly lower for larger system sizes. However it appears that the amplitudes of the peaks are also sensitive to the discretization of the momenta $k$ and how close the values are to the exact value $k = \frac{\pi}{2}, \pi, \frac{3 \pi}{2}$. This effect might be even more important than the system size.

We also observe that the width of the peaks changes with the system size. This is clearly a consequence of the system size $L$, and the discretization effect plays a minor role in this case.
The peaks at $k = \pi$ are less wide with larger values of $L$, however they appear to saturate to a finite width, see Fig.~\ref{fig:FiniteSizeEffectStructure}~a)-c).
On the other hand, the width of the peaks at half-integer values of $\pi$ do not appear to have strong system size dependence.
Such observation supports our claim that the integer peaks indicate a Mott state whereas the half-integer peaks signal short-range correlations and represent the Luttinger liquid of confined mesons.

\begin{figure}[t]
\centering
\epsfig{file=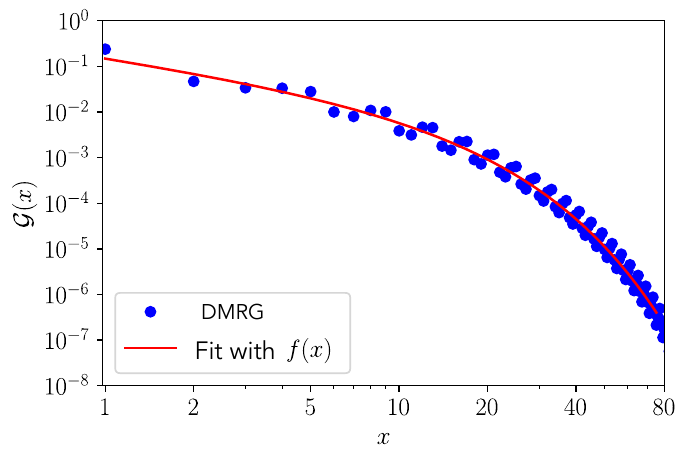, width=0.48\textwidth}
\caption{Green's function, Eq.~\eqref{eq:Greens}, at $V = 2h = 4t$ calculated using \textsc{SyTen} (blue points). The solid line is the result of fitting the DMRG data with Eq.~\eqref{eq:FullFitLog}.}
\label{fig:GreensPlasmaTypical}
\end{figure}

\section{Details on Green's function fits}\label{GreensFitsDetails}

\emph{Full fits.--}
The fit with Eq.~\eqref{eq:FullFit} was actually performed by taking the logarithm of the data and the distance $x$ in order to capture the low values of the exponential decay better. This means that we also had to use the logarithmic version of Eq.~\eqref{eq:FullFit} which equals to:
\begin{equation}
    f_f' = A_f' -\alpha_f x' -\beta_f e^{x'},
    \label{eq:FullFitLog}
\end{equation}
where $f_f' = \log{(f_f)}$ and $x' = \log{(x)}$. A typical fit can be seen in Fig.~\ref{fig:GreensPlasmaTypical}

\begin{figure}[t!]
\centering
\epsfig{file=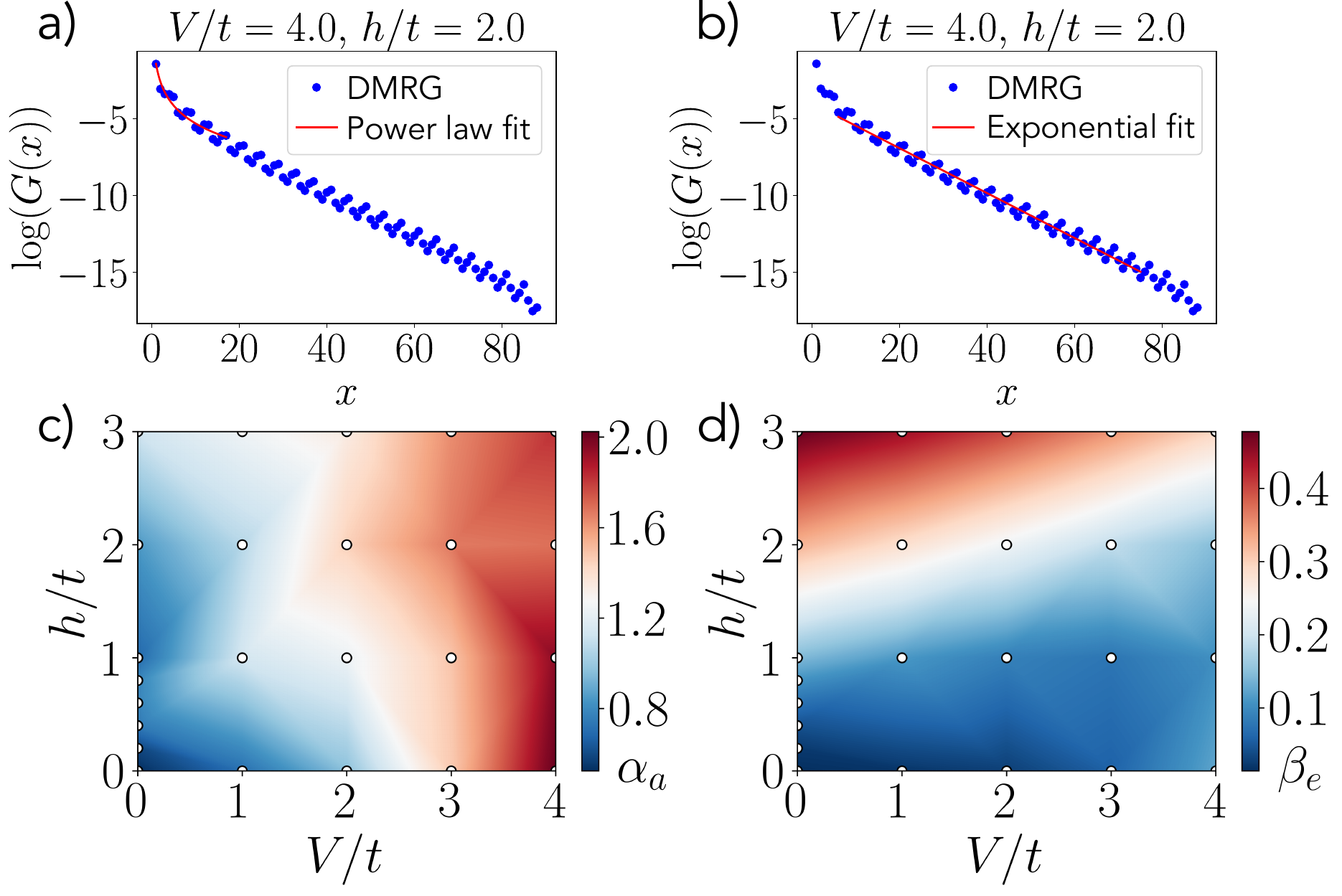, width=0.45\textwidth}
\caption{DMRG data of the \Zt invariant Green's function for $V = 2h = 4t$ and the fit with the power law function Eq.~\eqref{eq:AlgebraicFit} a) and exponential function Eq.~\eqref{eq:ExponentialFit} b). a) The first few data points nicely agree with the algebraically decaying fit function Eq.~\eqref{eq:AlgebraicFit} (red). b) On the other hand, for longer distances, the strong exponential decay agrees nicely with the exponentially decaying fit function Eq.~\eqref{eq:ExponentialFit} (red), thus signaling confinement of particle pairs into tight dimers. c) Heat map of the parameter $\alpha_a$, indicating the strength of the power law decay, for different parameter values $h, V$. d) Heat map of the parameter $\beta_e$ indicating the strength of the exponential decay for different parameter values $h, V$. Note that all fits and plots are in the log-linear plot.}
\label{fig:GreensDiagram}
\end{figure}

\emph{Separate fits.--}
We also address the power law and exponential decay separately by fitting the data with algebraic and exponential function respectively. For the intermediate distances we were able to fit the data with an algebraically decaying function
\begin{equation}
    f_{a} = A_a x^{-\alpha_a},
    \label{eq:AlgebraicFit}
\end{equation}
where $A_a$, is a free fitting parameters and $\alpha_a$ is reminiscent of the Luttinger liquid parameter. For the long distance behaviour, on the other hand, we fit our data with an exponentially decaying function
\begin{equation}
    f_{e} = A_e e^{-\beta_e x},
    \label{eq:ExponentialFit}
\end{equation}
where similarly as before $A_e$ is the free fitting parameter and $\beta_e$ is a simple parameter which quantifies the strength of the exponential decay. In both cases we took the logarithm of the DMRG data and then performed the fit with the logarithmic versions of Eq.~\eqref{eq:AlgebraicFit} and Eq.~\eqref{eq:ExponentialFit}.

Typical fit results can be seen in Fig.~\ref{fig:GreensDiagram} a) and b) for $2h = V = 4t $ where a clear exponential tail can be seen. The fit with Eq.~\eqref{eq:AlgebraicFit} nicely coincides with the first few data points on the intermediate length-scales. On the other hand the fit with the exponentially decaying function, Eq.~\eqref{eq:ExponentialFit} nicely captures the data points on the longer-length scale.
The extracted results of $\alpha_a$ and $\beta_e$ for different parameter values $h,V$ are presented in Fig.~\ref{fig:GreensDiagram} c) and d). The results are qualitative similar to those in the main text where the full function was fitted.

An important quantity is also the Luttinger liquid parameter $K$, which appears in many long-distance correlation functions. For free fermions the Luttinger liquid parameter is equal to one, $K = 1$, while for repulsive interactions it takes values smaller than one $K<1$ \cite{Giamarchi2004}. At the Mott transition point the Luttinger parameter equals to $K = 1/2$ and drops to $K = 1/4$ in the crystalline Mott state, for details see Ref.~\cite{Giamarchi2004}.

The parameter $\alpha_a = \frac{1}{2K^*}$ could be related to an inverse parameter $K^*$ which somewhat resembles the Luttinger liquid parameter. Our values for $h = V =0$ equals to $K^* = 0.95 \pm 0.02$ which is close to the Luttinger liquid parameter for free fermions $K = 1$ \cite{Giamarchi2004}.

On the other hand, the $K^*$ drops to the values close to $K^* = 1/4$ for $V \geq 2t$ which is once again similar to the behaviour of the Luttinger parameter at the transition point of the Mott state \cite{Giamarchi2004}. $K^*$ also decreases for $h > 0$ however in such regime the fits with the algebraic function became more challenging, since the increasing value of $h$ results in stronger exponential decay. This is reflected in $\beta_e$  which increases significantly with increasing values of $h$, see Fig.~\ref{fig:GreensDiagram} d). The increase of the parameter $\beta_e$ is offset by increased value of $V$. The line front of constant $\beta_e$ appears to be proportional to $2h = V$. This is once again in agreement with our numerical data of the charge gap and the structure factor, as well as our analytical arguments.

The chain lengths used for Green's function calculations were $L = 120$ and the starting point of our lattice was chosen to be away from the edge of the chain at $i_0 = 30$ in order to avoid strong edge effects. The distance $x$ is thus to be understood as $x = j - i_0$ where $j > i_0$.



%

\end{document}